\documentclass{aa}  

\usepackage{graphicx}
\usepackage{txfonts}
\usepackage{booktabs}
\usepackage{natbib}
\usepackage[normalem]{ulem}
\usepackage{siunitx}

\def\hd20{HD~209458\,b}
\def\gj34{GJ~3470\,b}
\def\hat11{HAT-P-11\,b}

\begin{document} 

   \title{Transmission spectroscopy and Rossiter-McLaughlin measurements of the young Neptune orbiting AU Mic}
   \titlerunning{AuMic with ESPRESSO}

  \author{E. Palle\inst{\ref{iiac},\ref{iull}}
  \and M. Oshagh\inst{\ref{iiac},\ref{iull}}
  \and N. Casasayas-Barris\inst{\ref{iiac},\ref{iull}}
  \and T. Hirano\inst{\ref{inst:tokyo}}
  \and M. Stangret\inst{\ref{iiac},\ref{iull}}
  \and R. Luque\inst{\ref{iiac},\ref{iull}}
  \and J. Strachan\inst{\ref{qmary}}
  \and E. Gaidos\inst{\ref{inst:hawaii}}
  \and G. Anglada-Escude\inst{\ref{inst:ice},\ref{inst:ieec}}
  \and P. Plavchan\inst{\ref{inst:plav}}
  \and B. Addison\inst{\ref{inst:addi}}
 }

  \institute{
  	     \label{iiac} Instituto de Astrof\'isica de Canarias (IAC), E-38200 La Laguna, Tenerife, Spain
  	\and 
  	    \label{iull} Deptartamento de Astrof\'isica, Universidad de La Laguna (ULL), E-38206 La Laguna, Tenerife, Spain
    \and 
        \label{inst:tokyo} Department of Earth and Planetary Sciences, Tokyo Institute of Technology, 2-12-1 Ookayama, Meguro-ku, Tokyo 152-8551, Japan
     \and 
        \label{qmary} Queen Mary University
     \and 
        \label{inst:hawaii} Department of Earth Sciences, University of Hawai'i at M\"{a}noa, Honolulu, Hawaii 96822 USA 
     \and 
        \label{inst:ice}Institut de Ci\`encies de l’Espai (ICE, CSIC), Campus UAB, Can Magrans s/n, 08193 Bellaterra, Spain
    \and 
        \label{inst:ieec}Institut d’Estudis Espacials de Catalunya (IEEC), 08034 Barcelona, Spain
     \and 
        \label{inst:plav}   Department of Physics and Astronomy, George Mason University, 4400 University Drive, MSN3F3, Fairfax, VA 22030, USA.
     \and 
        \label{inst:addi}   University of Southern Queensland, West St, Darling Heights QLD 4350, Australia
}

  \date{Received dd February 2020 / Accepted dd Month 2020}

  \abstract
   {AU Mic~b is a Neptune size planet on a 8.47-day orbit around the nearest pre-main sequence ($\sim$20 Myr) star to the Sun, the bright (V=8.81) M dwarf AU Mic. The planet was preliminary detected in Doppler radial velocity time series and recently confirmed to be transiting with data from the TESS mission. AU Mic~b is likely to be cooling and contracting and might be accompanied by a second, more massive planet, in an outer orbit.   Here, we present the observations of the transit of AU Mic~b using ESPRESSO on the VLT. We obtained a high-resolution time series of spectra to measure the Rossiter-McLaughlin effect and constrain the spin-orbit alignment of the star and planet, and simultaneously attempt to retrieve the planet's atmospheric transmission spectrum.  These observations allow us to study for the first time the early phases of the dynamical evolution of young systems. We apply different methodologies to derive the spin-orbit angle of AU Mic~b, and all of them retrieve values consistent with the planet being aligned with the rotation plane of the star. We determine a conservative spin-orbit angle $\lambda$ value of $-2.96^{+10.44}_{-10.30}$, indicative that the formation and migration of the planets of the AU Mic system occurred within the disk. Unfortunately, and despite the large SNR of our measurements, the degree of stellar activity prevented us from detecting any features from the planetary atmosphere. In fact, our results suggest that transmission spectroscopy for recently formed planets around active young stars is going to remain very challenging, if at all possible, for the near future. 
   }

   \maketitle

\section{Introduction}
\label{sec:introduction}

Planetary physical and orbital properties are predicted to evolve over time as a result of external and internal forcing. In fact, the observed distributions of exoplanet sizes and semi-major axes suggest that many planets migrate from their initial birth locations. Migration can occur by torques from primordial disks or scattering by a second planet and circularization by tides \citep{Lin1996,Kley2012}. Planets with gas envelopes will also cool and contract as they radiate their initial entropy of formation; close-in planets can also loose gas due to elevated irradiation by the active young host star. Detection and characterization of planets in their early formation stages ($<$1 Gyr) are essential to test models of these phenomena \citep{Baruteau2016}.

Detection of planets around young stars, however, is exceptionally challenging. Direct imaging methods are sensitive only to very massive planets at large separations and provide planet radii, but not masses.  Doppler radial velocity methods may be poised to overcome elevated stellar noise (\emph{jitter}) among young stars (e.g. \citet{Prato2008}) to detect and measure masses of close-in giant planets, but most of these will not transit and hence planet radii cannot be determined. Photometry from the Kepler space telescope has revealed transiting planets in 10-800 Myr-old clusters (e.g. \cite{Mann2016}) but clusters are distant, and most host stars are too faint to measure masses via RV and/or obtain detailed follow-up observations. On the other hand, members of \emph{young moving groups} (dispersed associations with similar space motions, abundances and ages) include exceptionally young (20-300 Myr) nearby ($<$50 pc) stars, many of which are pre-main sequence M dwarfs.  The closest such star known is AU Microscopii, an M-type member of the $\sim$20 Myr-old $\beta$ Pictoris Moving Group that is only at a distance of 9.8 pc.  AU~Mic has an edge-on debris disk with evidence of ongoing planet formation \cite{ Boccaletti2018, Daley2019}. Radial velocities of AU~Mic obtained  with both optical and infrared spectrographs (Plavchan et al., 2020) suggest the presence of one or more giant planets with orbital periods between 10 and 60 days. 

The TESS mission \citep{Ricker2014} observed AU~Mic for 27 days during Sector 1 of its prime survey. Three transits were identified by visual inspection, and their high significance verified against different models for stellar noise and false positives (Plavchan et al., 2020). Two transits with similar depth (0.3\%) and duration (3 hrs) were assigned to a candidate \emph{b} planet with an orbital period of 8.46 days and a radius of $\sim$ 0.4 R$_{\rm jup}$. Ground-based radial velocity measurements place the object's mass to be < 1.8 $M_J$, and bulk density of 0.25$\pm$0.15 g/cm$^{-3}$), indicating a low density planet (possibly still contracting) with a large scale-height. A third transit in the TESS light curve is within 1-$\sigma$ of the predicted time for a stronger radial velocity planet candidate, with a period of 30.7 days.

\begin{figure*}[h]
	\centering
	\includegraphics[width=6in]{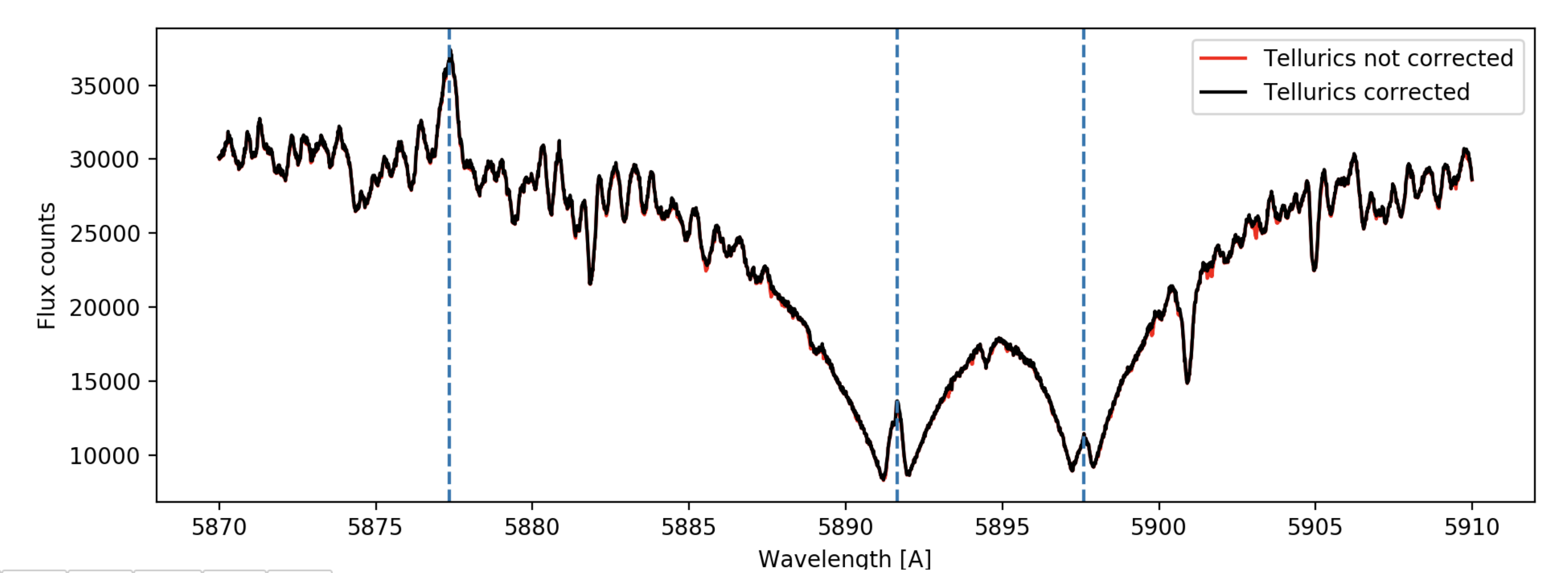}
	\caption{An out-of-transit spectrum of AU Mic, zooming on the spectral region containing the Na I doublet at 5891.7 and 5897.6 \AA, and the He I line at 5877.2 \AA. The broken lines indicate the central positions of these three lines. The red spectrum is the raw spectrum after DRS data subtraction, and the black spectrum results after the Molecfit telluric correction. }
	\label{fig:rawspec}
\end{figure*}

AU Mic now stands out as the closest system of transiting planets around a pre-main sequence star. This permits precise measurements of orbit, mass and radius which will allow us to rigorously test models of planet migration and evolution. Moreover, it has a debris disk which presumably marks the orientation of the former protoplanetary disk. The star's rotational equator, its disk, and the orbits of planets \emph{b} and \emph{c} seem highly aligned with our line of sight, but they might not be aligned with each other. Formation and migration of the planets within the original protoplanetary disk of AU~Mic should leave the planets on orbits with low inclinations with respect to the stellar rotation axis, while incipient or past scattering could leave one or more planets on orbits that are highly inclined to both the stellar equator and the disk. The Rossiter-McLaughlin (RM) effect and doppler tomography are an excellent techniques to infer the obliquity/inclination of fast-rotating stars like AU Mic. Here we present spectroscopic observations of one transit of the inner planet, AU Mic~b, to constrain the projected angle of the planet's orbit with respect to the star's rotation axis, and perform an initial exploration of possible chemical species (atomic and molecular) in its atmosphere.

\section{Observations and Data Reduction}
\label{sec:Observations}

One transit of AU~Mic~b was observed using the high precision RV spectrograph ESPRESSO at the VLT in its standard setup (400--780nm, 1 UT - HR, standard calibration), during the night of August 7th 2019. Data were taken continuously from 3:24UT to 9:23 UT, with an exposure time of 200 s. A total of 88 on-target spectra were taken, of which 37 spectra were taken out of transit, while another 49 spectra were taken in-transit, over the $\sim$3.5 hours that the transit lasted. This high-cadence sampling is in principle an advantage, given that the star is active and variable, for the modeling and removal of line profile changes. During the observations, the airmass varied from 1.03 to 2.37, passing through a minimum of 1.007. Weather conditions were clear. The averaged S/N of the spectra per pixel, measured at order 104 (near the Na I doublet), was 93.9.



The spectra were extracted and calibrated using the standard ESO-Data Reduction Software (DRS). According to the ESPRESSO exposure time calculator (ETC), the resulting individual RV measurements, considering photon noise and the instrumental floor, should have been better than 1~m/s. However, the extracted RVs have a median internal precision of 3.7 m/s, resulting from the broadening of the spectral feature of AU Mic due to rotation and stellar activity. Note that Plavchan et al (2020) report AU Mic to be very active relative to main-sequence dwarfs, and find RV peak-to-peak variations, using HARPS spectrograph data, in excess of 400 m/s primarily due to the rotational modulation of stellar activity.

We also used a second approach to retrieve the RV values during the observations. We applied SERVAL \citep{SERVAL} to the 2D ESPRESSO spectra produced by the DRS pipeline. SERVAL produces high-precision differential Doppler observations by computing them relative to a high SNR template (co-added outside transit spectra). Using SERVAL we achieve a median internal precision of 4.3 m/s. SERVAL also produces several activity indicators such as the differential line width (DLW) and chromaticitity index (CRX), useful for our analysis and interpretation.

\section{The Rossiter-McLaughlin effect}
\label{sec:rm}

\begin{figure}
	\centering
	\includegraphics[width=\columnwidth]{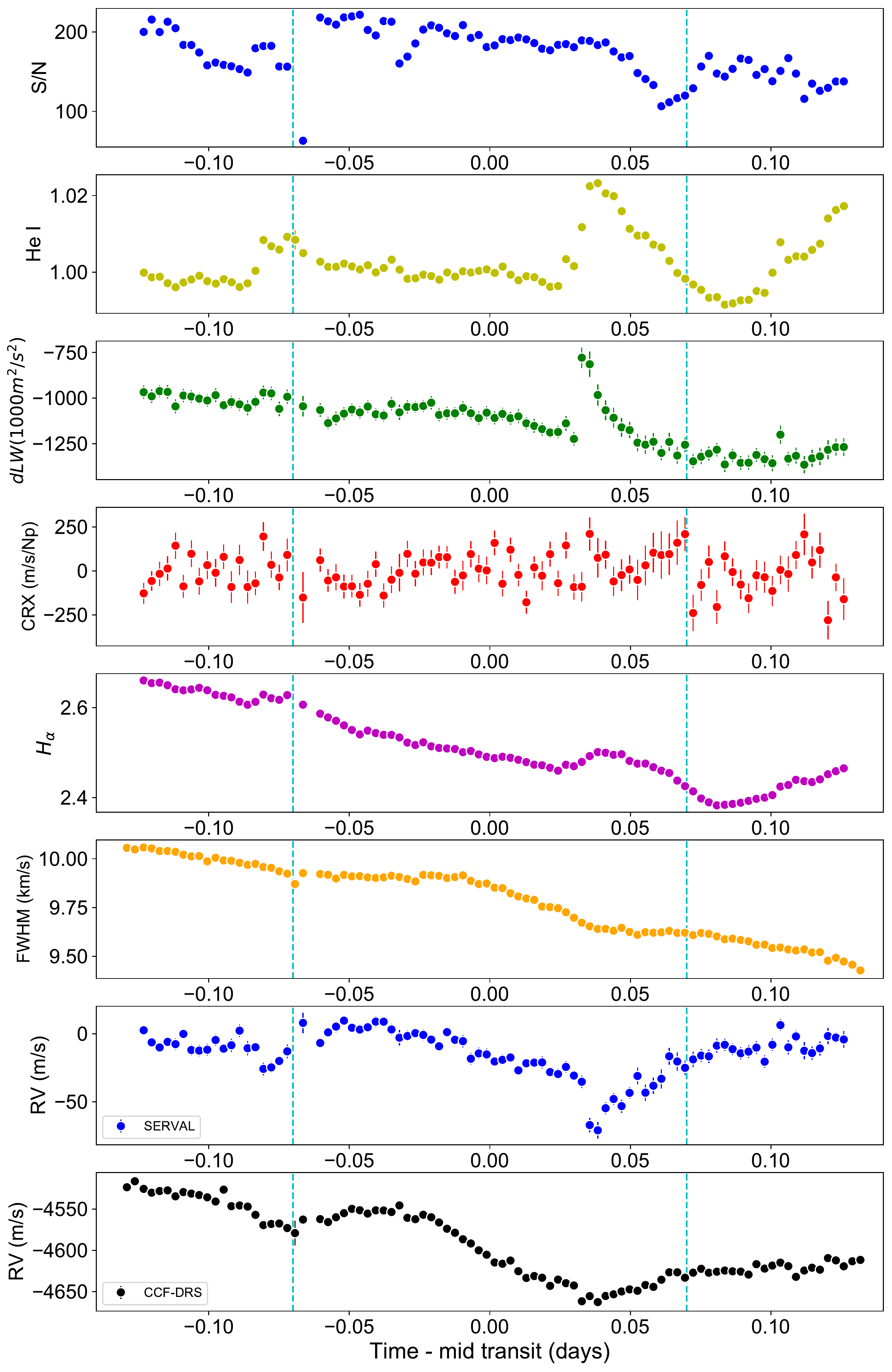}
	\caption{From top to bottom: The S/N of each of the individual spectra per pixel; The light curve of the chromospheric He I line at 5877.2 \AA. The spectral line was integrated over a spectral range of 1.5\AA-wide; the dLW and CRX indices derived using SERVAL; The light curve of the $H_{\alpha}$ line; The FWHM of CCF as estimated from DRS; The radial velocities derived from SERVAL; and the radial velocities derived from DRS. The vertical dashed cyan lines mark the predicted ingress and egress of transit of AU~Mic~b.}
	\label{fig:indicators}
\end{figure}



The transit of an exoplanet in front of its rotating host star generates a detectable RV signal, as the planet blocks the corresponding rotational signal of a portion of the stellar disk. This regional signal is removed from the integration of the velocity over the entire star, known as the RM effect \citep{Holt-1893, Rossiter-24, McLaughlin-24}. The RM observation is a powerful and efficient technique for estimating the spin-orbit angle of exoplanetary systems (see \citet{Triaud-18} and references therein for a comprehensive review). 

\subsection{Transit time series and stellar activity}

The rotation period of AU Mic (4.8 days) is long compared to the transit duration. Except for the occurrence of a flare (easy to identify in chromospheric emission lines) the changes in the spectrum due to photospheric features should be smooth during the $\sim$6~hr that our observations lasted. Therefore, any change in RV due to rotational modulation can be removed by interpolating the out-of-transit line-profile measurements. 

The ESPRESSO times series, however, shows strong evidences of flares and stellar activity during the observations. Figure~\ref{fig:rawspec} shows a cut of an out-of-transit spectrum of AU Mic. The selected spectral region contains the Na I doublet (5891.7 and 5897.6 \AA) and the He I line at 5877.2 \AA. It can be readily identified in Figure~\ref{fig:rawspec}  that AU Mic presents strong signatures of stellar activity. The total flux on the Na I doublet has two components; a chromospheric emission core is superimposed on the much broader and deeper photospheric absorption line \citep{Walkowicz2008}. In comparison, for the stellar chromospheric emission line of He I, only an emission feature is visible. On top of that, these emission features vary with time during the transit of AU~Mic~b, making it challenging to retrieve the orbital and atmospheric planet properties. Interestingly, the emission lines have similar shapes, suggesting co-localization of the emission. 

In Figure~\ref{fig:indicators} we plot the time evolution of several parameters that need to be taken into account for the analysis of AU~Mic~b's transit. In particular, the SNR of each of the individual spectra and the transit light curve of the chromospheric He I line (see Section~\ref{sec:trans}) which we use as a proxy for stellar flares, are useful indicators to weight the value of each observations for the RM determination. 

\subsection{RM calculation}

There are two main techniques to measure RVs during the transit of an exoplanet and obtain the RM signal. One approach relies on the template matching of the observed spectra \citep{Butler-96, Zechmeister-18}, and the other one is based on a Gaussian fit to the cross-correlation function (CCF) of the observed spectra with a binary mask \citep{Pepe-02}.  Each approach leads to a different shape of RM signal as was demonstrated theoretically in \citet{Boue-12}. 

To model the RM observation obtained using the CCF- DRS pipeline, we use the publicly available code \texttt{ARoME} \citep{Boue-12} which is optimized to model the RM signal extracted through CCF-based approach. To model the RM observations obtained from template matching using the SERVAL pipeline, we use the model based on \citet{Ohta-05} approach,  which is optimized to retrieve the RM signal from template matching. This model is implemented in the \texttt{PyAstronomy} python package. 

For both analysis, we considered a second order polynomial trend to interpolate the out-of-transit RV data between ingress and egress. This trend is present in both RM observations either from CCF-DRS or SERVAL, however, with different shape and strength. This trend could have several contributors, either from the planet's Keplerian orbit, any possible systematics in the data, and also the stellar activity induced RV \citep{Oshagh-18}. Thus, during our fitting procedure (either with \texttt{PyAstronomy} or \texttt{ARoMe}) we consider the spin-orbit angle $\lambda$, projected stellar rotational velocity ($v sin(i)$), mid-transit time ($T_0$), limb darkening coefficients, and parameters of the second order polynomials as our free parameters. The rest of the parameters required in the models are fixed to their reported values in the literature ($P_{orb}$= 8.46321 days, $R_{p}/R_{\star}$=0.0514, $a/R_{\star}$=19.1, and impact parameter (b)=0.16; Plavchan et al, 2020). Based on the time series shown in Figure~\ref{fig:indicators}, and again for both analysis, one data point with very low S/N value just prior to the transit ingress was discarded, and several data points affected by a strong flare during the egress (starting at 0.29 days from the mid transit till the end of transit) were given a low weight during the fitting procedure.


The best fit parameters and associated uncertainties in our fitting procedure are derived using a Markov chain Monte Carlo (MCMC) analysis, using the affine invariant ensemble sampler \texttt{emcee} \citep{Foreman-Mackey-13}. The prior on $v sin(i)$ and $T_0$ are controlled by Gaussian priors centered on the reported value in the literature and width according to the reported uncertainties, and the prior on spin-orbit angle is also controlled by a uniform (uninformative) prior between -180 and +180 degrees. These priors are also listed in Table~\ref{tab:fitted RM values}. We randomly initiated the initial values for our free parameters for 30 MCMC chains inside the prior distributions. For each chain we used a burn-in phase of 500 steps, and then again sampled the chains for 5000 steps. Thus, the results concatenated to produce 150000 steps. We determined the best fitted values by calculating the median values of the posterior distributions for each parameters, based on the fact that the posterior distributions were Gaussian. 

\begin{figure*}
	\centering
	\includegraphics[width=\columnwidth]{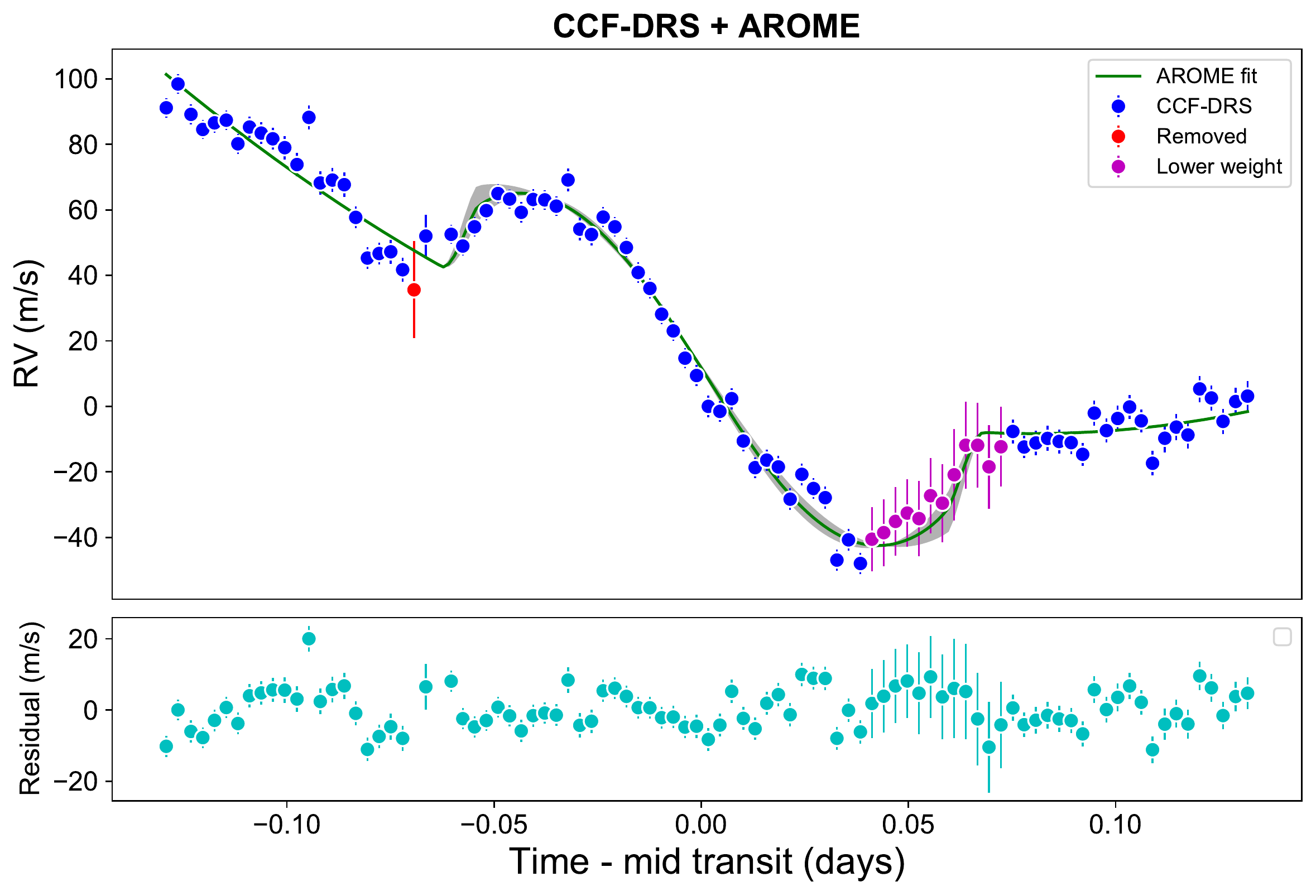}
	\includegraphics[width=\columnwidth]{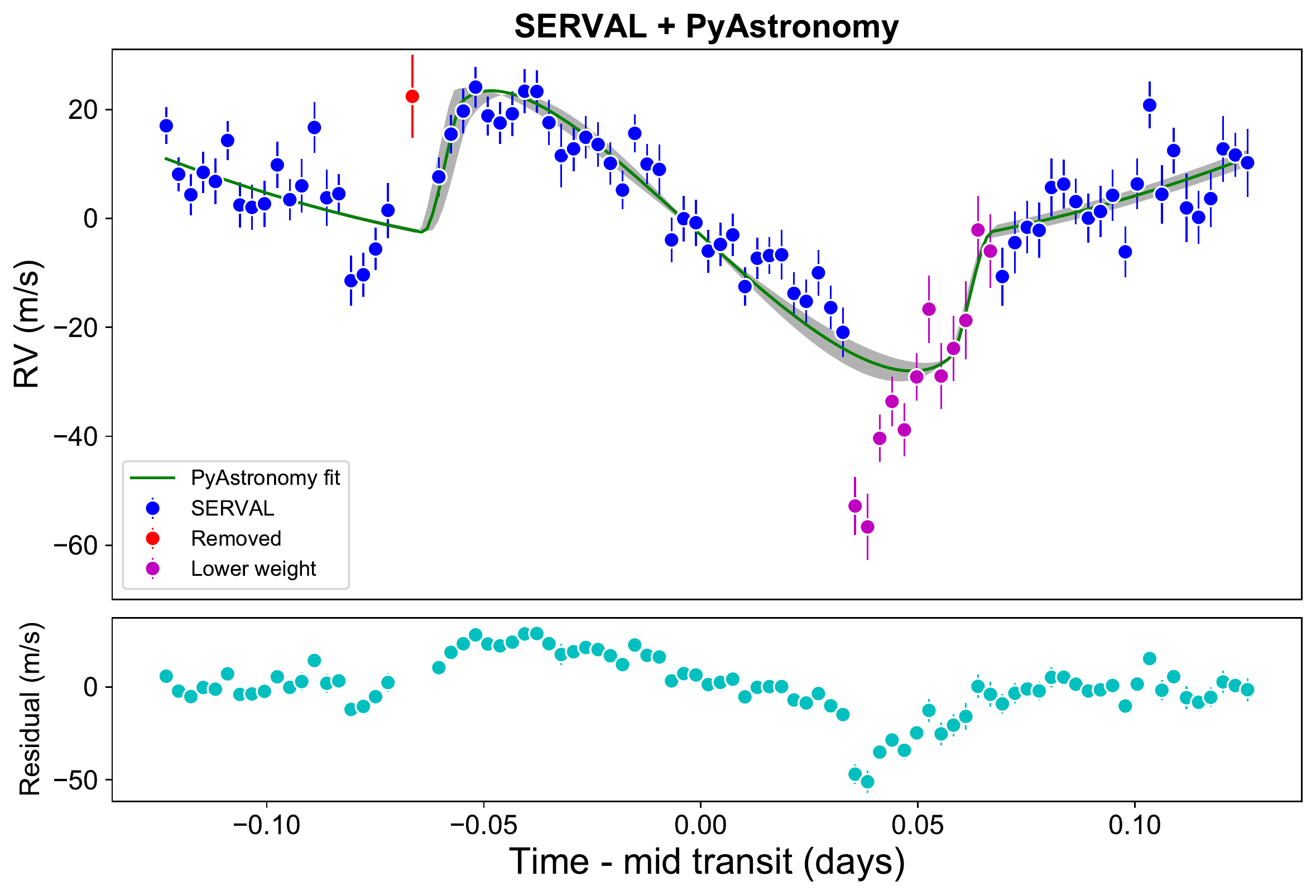}
	\includegraphics[width=\columnwidth]{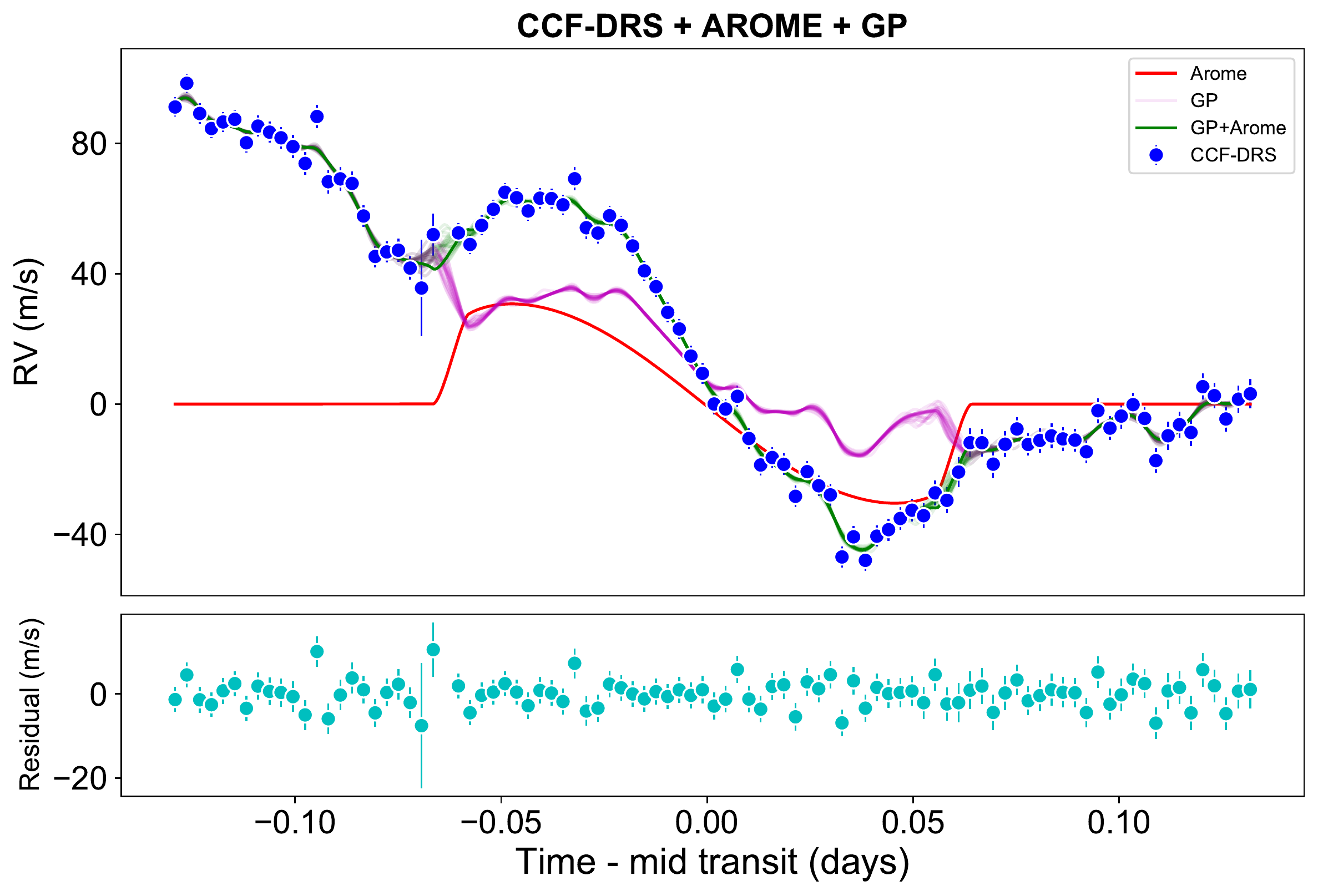}
	\includegraphics[width=\columnwidth]{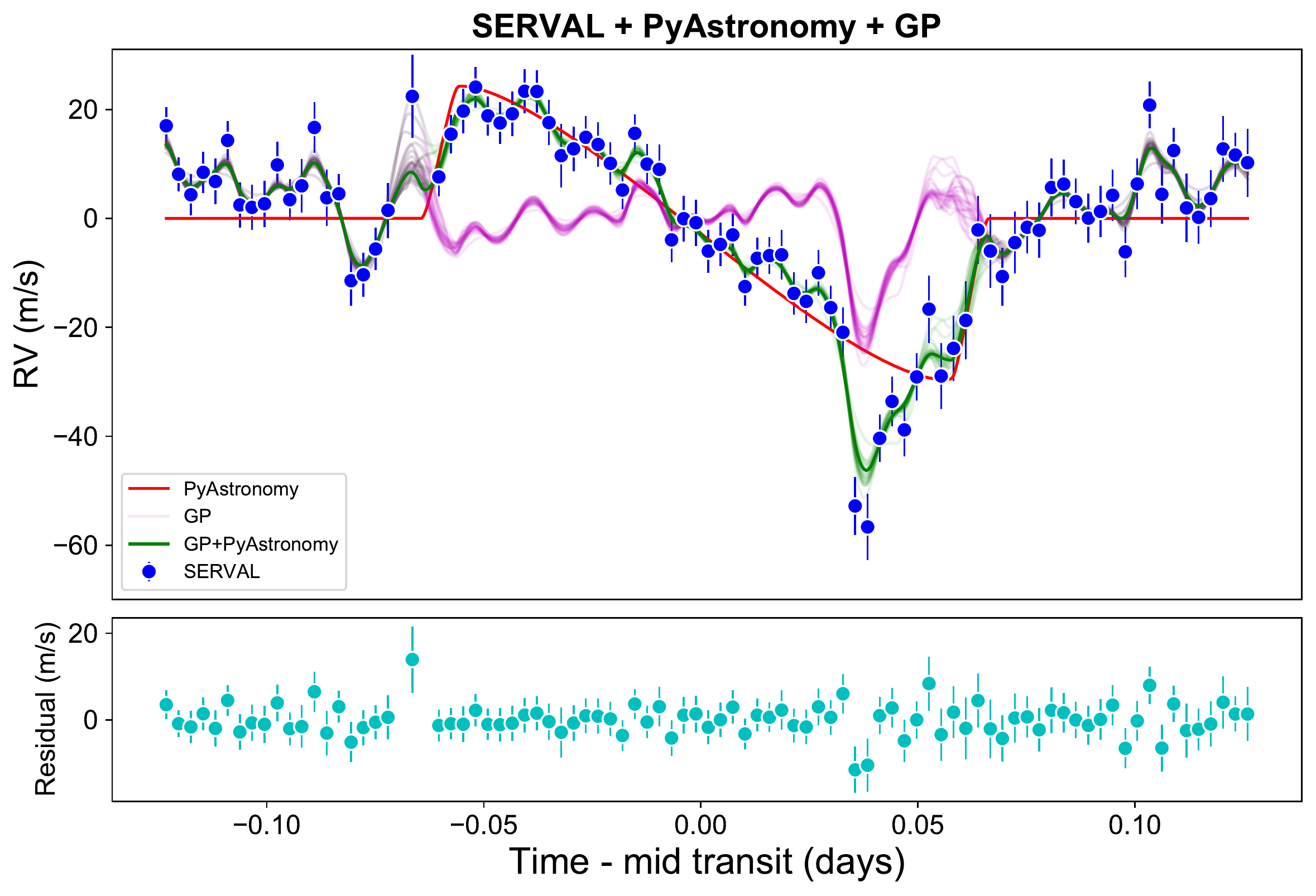}
	\caption{Radial velocity time series derived from the DRS software (left columns) and SERVAL (right columns). On the top panels the best fit model to the RM is shown using the \texttt{ARoME} and \texttt{PyAstronomy} models, respectively. The red point is discarded in our analysis due to anomalously low SNR, while the magenta points are given a lower weight in the RM fits due to being affected by a stellar flare.
	In the lower panels, the data are the same but the fits incorporate the GP modelling. The different components of each best fit model are plotted in different colors and marked in the legend.}
	\label{fig:rmeffect}
\end{figure*}

The posterior distributions for both analysis are given in Figures~\ref{fig:corner1} and ~\ref{fig:corner2}, and the best fitted models (both \texttt{PyAstronomy} and \texttt{ARoME}) analysis and RM observations (both DRS-CCF and SERVAL) are shown in Figure~\ref{fig:rmeffect}. The DRS-CCF and \texttt{ARoME} analysis suggest that the planet is aligned (spin-orbit angle =0), however, it overestimates the $v sin(i)$ value in comparison to that from spectral line analysis as reported in Plavchan et al, 2020. On the other hand the SERVAL and \texttt{PyAstronomy} analysis suggest a slightly misaligned planet (spin-orbit angle =-9 degree) but consistent with zero within the error bar, and also estimated the $v sin(i)$ to be consistent with the reported value from spectral analysis. We report all these value in Table~\ref{tab:fitted RM values}.

\begin{table*}[ht]
	\caption{Best fit derived values for the DRS+\texttt{ARoME} and SERVAL+\texttt{PyAstronomy} fits, with and without the use of Gaussian processes.}
	\centering
	\begin{tabular}{cccccc}
    \hline\hline 
		Parameter & Prior & ARoME & PyAstronomy & ARoME+GP & PyAstronomy+GP \\
		\hline
		$\lambda^{\circ}$ & $\mathcal{U}(-180;180)$&  $0.43^{+1.87}_{-2.04}$&  $-10.87^{+5.55}_{-5.05}$ & $- 2.96^{+10.44}_{-10.30}$ & $6.61^{+11.62}_{-12.35}$\\
		$v \sin i (km\,s^{-1})$ &  $\mathcal{N}(8.7; 2)$& $15.66^{+0.42}_{-0.93}$ & $12.51^{+0.51}_{-0.54}$ & $11.15^{+3.71}_{-3.42}$ & $10.42^{+1.52}_{-1.50}$ \\
		$\epsilon$ &   $\mathcal{N}(0.62; 0.1)$& - & 0.72 & - & 0.39\\
		$u_{a}$ &   $\mathcal{N}(0.47; 0.1)$& 0.57 & - & 0.35 &-\\
		$u_{b}$ &    $\mathcal{N}(0.30; 0.1)$& 0.20 & - & 0.16 & -\\
		$A_{GP}(m\,s^{-1})$ & $\mathcal{U}(0.1;100)$& - & - & 27.11& 8.10\\
		$\tau_{GP} (days)$ & $\mathcal{N}(0.018; 0.01)$& - & - & 0.025 & 0.007\\
		RMS Residual $(m\,s^{-1})$ &-& 6.10 & 9.78 & 3.80 & 3.42\\
		$ln\,Z$ &-& $-369.76 \pm 0.20$ & $-307.76 \pm  0.08$ & $-326.18 \pm 0.11$ & $-293.21 \pm  0.04$\\

		\hline

	\end{tabular}
	\label{tab:fitted RM values}
	\begin{flushleft} 
\textbf{Notes}:  $\mathcal{U}(a;b)$ is uniform prior with lower and upper
limits of a and b, $\mathcal{N}(\mu; \sigma)$ is a normal distribution with mean $\mu$ and width $\sigma$. $T_{0}$ mean reported are the first transit time for each planet.

\end{flushleft}
\end{table*}

\subsection{Modeling RM with additional Gaussian Process} \label{sec:RM-GP}

Gaussian process is a general framework for modeling correlated noise \citep{Rasmussen-06}, and it has been shown its power and advantages in modelling and mitigating the stellar activity noise in RV observations \citep[e.g.,][]{Haywood-14, Faria-16}, and also in photometric transit observations \citep[e.g.,][]{Aigrain-16, Serrano-18}, which assisted in detecting small-sized planetary signals embedded in the stellar activity noise.

Since our RM observations (either DRS or SERVAL) are clearly affected by the stellar activity noise (such as stellar spots occultations by the transiting planet and flares), we decided to incorporate GP to our RM modeling, in order to perform more robust fit and obtain more accurate estimates. We used the new implementation of GP in \texttt{celerite} package \citep{Foreman-Mackey-17}, since some of the \texttt{celerite} kernels are well suited to describe different forms of stellar activity noises. To model the stellar noise in our data set we selected the covariance as a Matern-3/2 Kernel. In order to train our GP, we first fit this GP model to the differential line
width (DLW) and He I light curve, which are the most affected observables by the flares occurring during our observation. This fitted GP allows us to have a better estimate on the prior time scale of the noise in our RV measurements.
Here, we fit again our RM observations, however, this time we modeled the observed RMs (either from DRS or SERVAL) as the sum of the mean model and the noise. The mean model is the RM model (either \texttt{AROME} or \texttt{PyASTRONOPMY}, depending on which RM observations), and the noise was modeled as a Gaussian process with Matern-3/2 covariance Kernel. The posterior samples for our model were obtained through MCMC using emcee \citep{Foreman-Mackey-13}. The prior on the RM models' parameters were controlled as in the previous section, and the prior on the GP time scale parameter was controlled by Gaussian priors centered on the fitted value from DLW and He I. And the prior on GP amplitude was controlled by a uniform (uninformative) prior between 0.1 to 100 m/s. These priors are also reported in Table~\ref{tab:fitted RM values}. The posterior distributions for both analysis are given in Figures~\ref{fig:corner3} and ~\ref{fig:corner4}, and the best fitted models (both \texttt{PyAstronomy}+GP and \texttt{ARoME}+GP) analysis and RM observations (both DRS-CCF and SERVAL) are shown in Figure~\ref{fig:rmeffect}. We report all the best fitted values in Table~\ref{tab:fitted RM values}.

The GP+RM model result in much better fit as indicated by significant decrease in the RMS of the residual (see reported RMS of residuals in  Table~\ref{tab:fitted RM values} . However, in order to perform a proper model comparison between RM only and RM+GP model we utilised \texttt{MultiNest} \citep{Feroz-09} via the \texttt{Dynesty} package \citep{Speagle-20}. Dynasty provide the Bayesian model log evidence (ln Z), and as suggested by \citet{Trotta-08}, we regard the difference between two models as strongly significant if their log-evidence differs by $\Delta ln Z > 5$. Our result show that there are strong evidence supporting the idea of fitting the observed RM with a GP given their noise (Table~\ref{tab:fitted RM values}).

The DRS-CCF and \texttt{ARoME+GP}, and SERVAL and \texttt{PyAstronomy+GP}  analysis both clearly demonstrated that the planet is prograde and to be aligned (within the uncertainties). However, both still overestimated the $v sin(i)$ value, although being closer to the reported value from spectral analysis as reported in Plavchan et al, 2020. The estimated $v sin(i)$, considering their uncertainties, are compatible with the photometric rotation period of the star reported in Plavchan et al, 2020.

\citet{Brown-17} performed a comparison study and showed that both models overestimate $v sin(i)$ in comparison to estimates obtained from spectral line broadening. They found that the overestimation can even reach as high as 5 km/s for fast rotating stars. However, they found the estimated spin-orbit angles were in strong agreement for both models. The overestimation of $v sin(i)$  could have variety of reasons. For instance, the RM signal can be affected by second-order effects such as the convective blueshift and granulation \citep{Shporer-11, Cegla-16}, stellar differential rotation \citep{Hirano-11, Hirano-14, Albrecht-12b,Cegla-16,Serrano-20}, microlensing effect due to the transiting planet's mass \citep{Oshagh-13}, impact of ringed exoplanet on RM signal \citep{Akinsanmi-18, deMooij-17}, occulted stellar active regions \citep{Oshagh-16,Oshagh-18}, non-occulted stellar active regions \citep{Boldt-20}, and also inaccurate estimations of stellar limb darkening \citep{Csizmadia2013,Yan-15}. Some of these effects will have minor impact on the RM shape and amplitude (such as microlensing and ring around the planet), however, some like the stellar active region occulation and stellar differential rotation could cause significant deformation of the RM shape and also alter its amplitude. None of mentioned second-order effects are considered neither in \texttt{ARoME} nor in \texttt{PyAstronomy}, and they could be responsible for the overestimation of $v sin(i)$. It also worth mentioning that the spectroscopic estimates are also model dependent (choice of macroturbulence, etc.), therefore they could be also underestimating the $v sin(i)$, however in depth investigation of the reason for our overestimated $v sin(i)$ beyond the scope of this study.

From all methodologies applied here, we choose a $\lambda$ value of $-2.96^{+10.44}_{-10.30}$, based on Arome and GP, as a conservative determination of the spin-orbit angle, based on the data at hand. 

\section{Doppler Tomography of AU Mic~b }
\label{sec:rm}

The Doppler tomography method \citep{Collier2010, Watson2019} measures changes in the rotationally-broadened stellar line profiles due to partial occultation by the planet. 
Since cross-correlation functions (CCFs) of the observed spectra against a template 
reflect the mean profile of the stellar lines, one can probe the instantaneous velocity 
field on the stellar disk from the time-varying cross-correlations. 
The fit to this differential profile allows for a more precise and model-independent
measurement of the free parameters: planet/star size radius ratio, limb-darkening, 
projected obliquity angle, impact parameter, and duration \citep{Strachan2017}.

\begin{figure*}
	\centering
	\includegraphics[width=18cm]{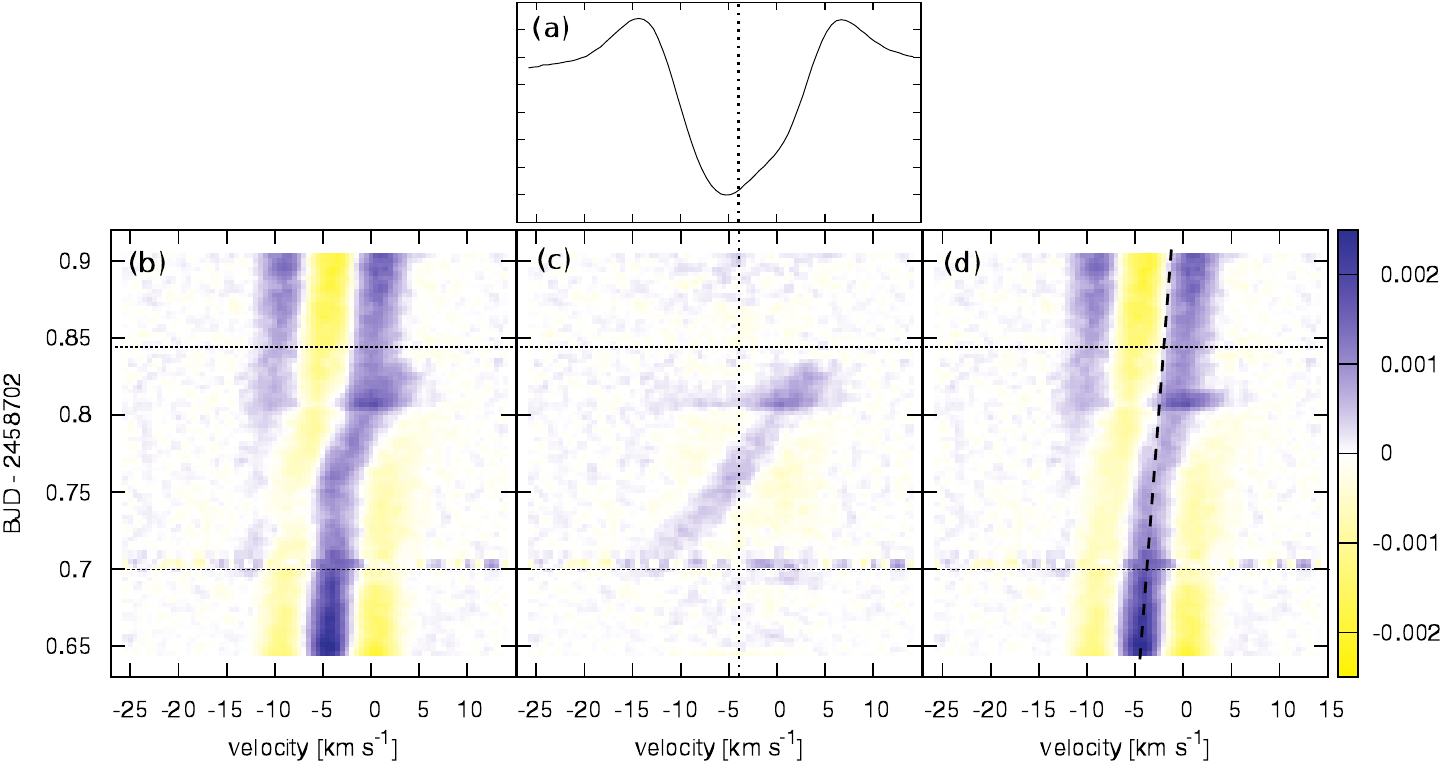}
	\caption{Sample CCF profile during the night (panel (a)) and 
	the residual (differential) CCF maps as a function of time (panels (b)-(d)). 
	Panel (b) illustrates the original residual CCF map after subtracting the 
	mean out-of-transit CCF. The low-frequency modulations were subtracted
	based on the out-of-transit variations in panel (c), in which the planet
	shadow is clearly seen. Panel (d) shows the residual map after the observed 
	planet shadow is removed. The vertical dotted lines in panel (a) and (c) are
	the approximate CCF center, representing the systemic RV of AU Mic. 
	The near-vertical dashed line in panel (d) draws the maximum shift
	of the CCF bump by a spot-like feature on the stellar surface 
	($\approx 3$ km s$^{-1}$ over 0.25 day). 
	}
	\label{fig:DT1}
\end{figure*}

For an independent measurement of the stellar obliquity, we analyzed the time sequence of the CCFs returned by the DRS pipeline for the ESPRESSO spectra (see Figure~\ref{fig:DT1}a for an observed CCF). To visualize the line-profile ``variation", we computed the residual cross-correlation 
map as a function of time; in doing so, we first normalized the CCF for each
frame, and combined the CCFs for frames taken outside the transit. 
We then subtracted the mean out-of-transit CCF from individual frames, 
yielding the residual CCF map in time. 

Figure~\ref{fig:DT1}(b) displays the resultant residual CCF map. The two horizontal dashed lines represent the expected transit ingress and egress times. The residual map exhibits an unexpectedly large modulations throughout our observations. 
This observed low-frequency modulation in the profile is not caused by AU Mic b's transit, and the reasons for the large profile modulation are not known. 

To remove this strong modulation, we applied a high-pass filtering in which we fitted the "out-of-transit" residual CCF by a quadratic function of time in each column of the residual CCF map. Interpolating the CCF variations during the transit, we subtracted the low-frequency modulation from the residual CCF map. Figure~\ref{fig:DT1}(c) shows the CCF map after the high-pass filtering. The CCF bump, representing the planet shadow of AU Mic b, is clearly seen in the map,  moving from the blue edge to the red edge of the original CCF (Figure~\ref{fig:DT1}(a)). The trajectory of the shadow implies a prograde orbit of the transiting planet. The residual CCF map also suggests a flaring event happening around  $\mathrm{BJD}=2458702.81$, which suddenly distorts the line profile for a relatively short interval. This is consistent with the large scatter in the observed RV data towards the
end of the transit (Figure~\ref{fig:rmeffect}).

\begin{figure}
	\centering
	\includegraphics[width=\columnwidth]{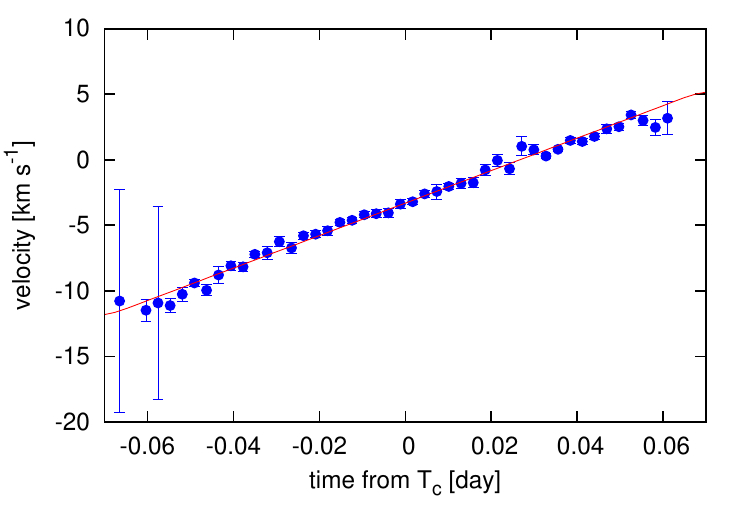}
	\caption{
	Observed velocity positions of the planet shadow in the CCF profile (blue points)
	as a function of time.
	The red solid line indicates the best-fit RM model. 
	}
	\label{fig:DT2}
\end{figure}

To estimate the stellar obliquity $\lambda$, as well as other system parameters, from the 
residual CCF map, we performed an analysis of the ``reloaded RM effect" \citep{Cegla2016}. 
In brief, we fitted the positions of the planet shadow during the transit using a Gaussian 
and estimated the instantaneous Doppler position of the planet in the CCF profile
for each frame (Figure~\ref{fig:DT2}).  
Then, using the MCMC code to model the RM effect \citep{2020ApJ...890L..27H}, 
we fitted the positions of 
the planet's shadow. The free adjustable parameters in the analysis are $\lambda$, $v\sin i$, 
and the CCF center $\gamma$, representing the peculiar RV of AU Mic. 
In many cases, $\gamma$ is determined precisely by fitting the mean CCF profile
outside the transit by Gaussian. In the case of AU Mic, however, the mean out-of-transit CCF of the star is highly distorted and asymmetric (Figure~\ref{fig:DT1}(a)), introducing an additional source of uncertainty in the CCF center. A fit to the mean out-of-transit CCF led to $\gamma=-4.16\pm0.46$ km s$^{-1}$, but the uncertainty here is likely underestimated due to the asymmetric CCF. 

The MCMC fit to the observed planet shadow positions without any priors resulted in a 
degeneracy among $\lambda$, $v\sin i$, and $\gamma$. Thus, we imposed a Gaussian prior on 
$v\sin i$ based on the spectroscopic value reported in the discovery paper 
(i.e., $v\sin i=8.7\pm 2$ km s$^{-1}$). 
This analysis resulted in $\lambda=0_{-19}^{+20}$ degree, which is consistent
with the RV analysis results in Section \ref{sec:RM-GP}. The best-fit model to the observed
planet shadow positions is drawn by the red solid line in Figure~\ref{fig:DT2}. 
To take into account the constraint on $\gamma$ from the mean CCF profile, we also 
attempted an MCMC analysis imposing priors on both $v\sin i$ and $\gamma$ 
($=-4.16\pm0.46$ km s$^{-1}$). This fit produced $\lambda=17_{-12}^{+9}$ degree, 
still compatible with the spin-orbit alignment of the system within $1.5\,\sigma$. 
Thus, we conclude that the RM fit and Doppler-tomography analysis yield fully 
consistent results, both independently supporting that the system has a low obliquity.

An unresolved issue is the large CCF modulation seen throughout the night of the transit. Figure~\ref{fig:DT1}(d) illustrates the residual CCF map after removing the planet shadow (panel c) from the original map. The CCF bump seen at $\approx -5$ km s$^{-1}$ at the beginning of the observations moves on the stellar disk by $\approx 5$ km s$^{-1}$ at the end of the run. For comparison, the AU~Mic~b planetary signal moves by $\approx 20$ km s$^{-1}$ during the shorter transit time. On the other hand, the rotation period of the star is estimated to be 4.863 days, and thus a star spot on the stellar surface should change its phase by $0.26/4.863=0.053$ during our observations. Assuming $v\sin i=8-9$ km s$^{-1}$, this phase shift translates to the maximum Doppler shift of $\approx 2.9$ km s$^{-1}$ around the center of the stellar disk. 
For reference, we plot in the same figure this maximum shift of a CCF feature by stellar
rotation. The observed shift looks consistent with the rotation up to around the mid-transit time, but suddenly it changed the Doppler position by a few km s$^{-1}$. 
At this point, we are not able to conclude that the observed modulation is an outcome
of a giant spot on the stellar surface or a redistribution of visible active regions during the stellar rotation; further Doppler monitoring is needed to understand the peculiar behaviour of the CCF profile.

\section{Searching for Atmospheric Signatures}
\label{sec:trans}

\subsection{Transmission spectroscopy}

\begin{figure}
	\centering
	\includegraphics[width=\columnwidth]{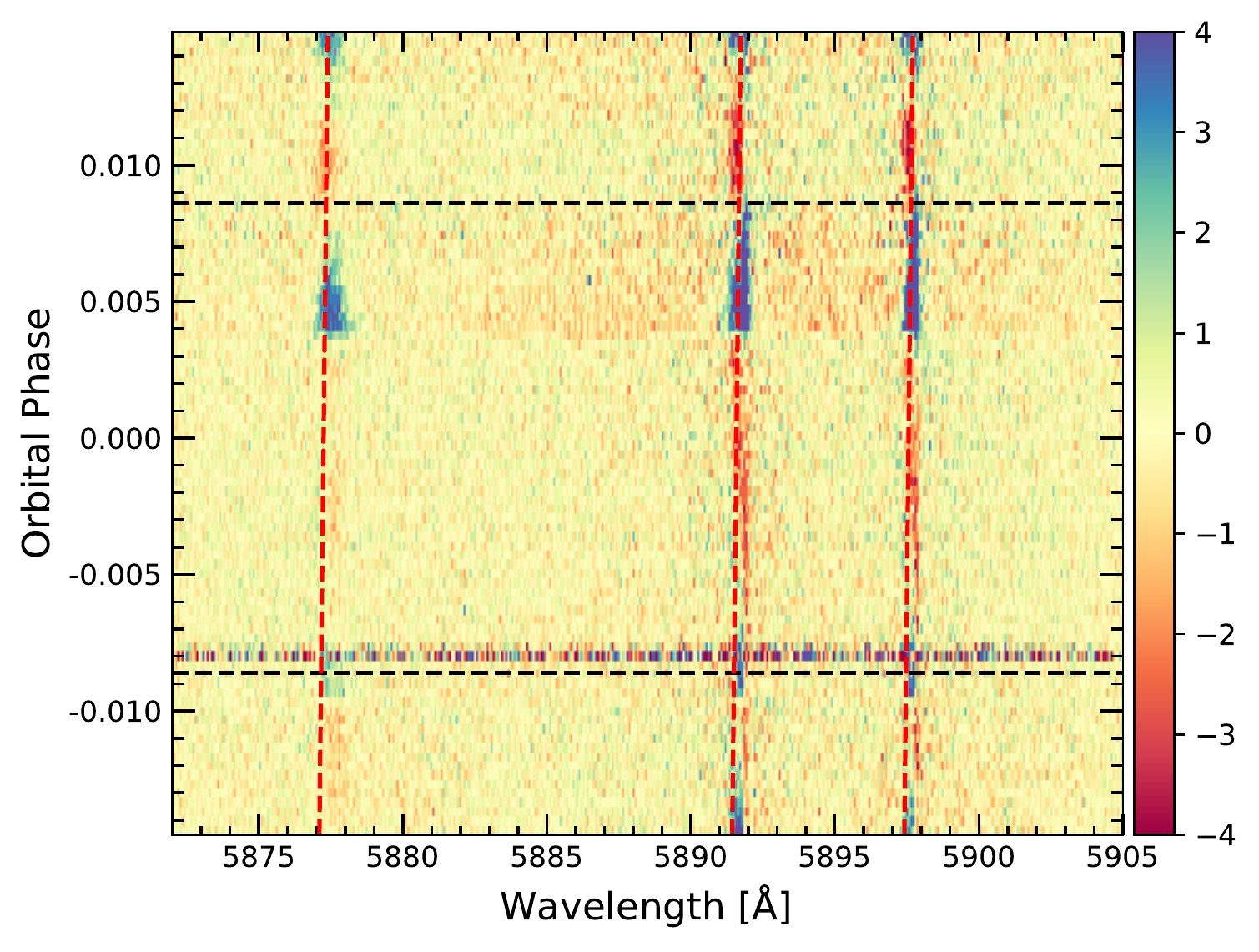}
	\caption{The observed 2D maps of residual spectra around the \ion{Na}{i} doublet, in the stellar rest frame.The horizontal dashed lines mark the T1 and T4 transit contacts. The red dashed lines mark the position of the \ion{Na}{i} and He I lines.}
	\label{fig:2danom}
\end{figure}

\begin{figure}
	\centering
	\includegraphics[width=\columnwidth]{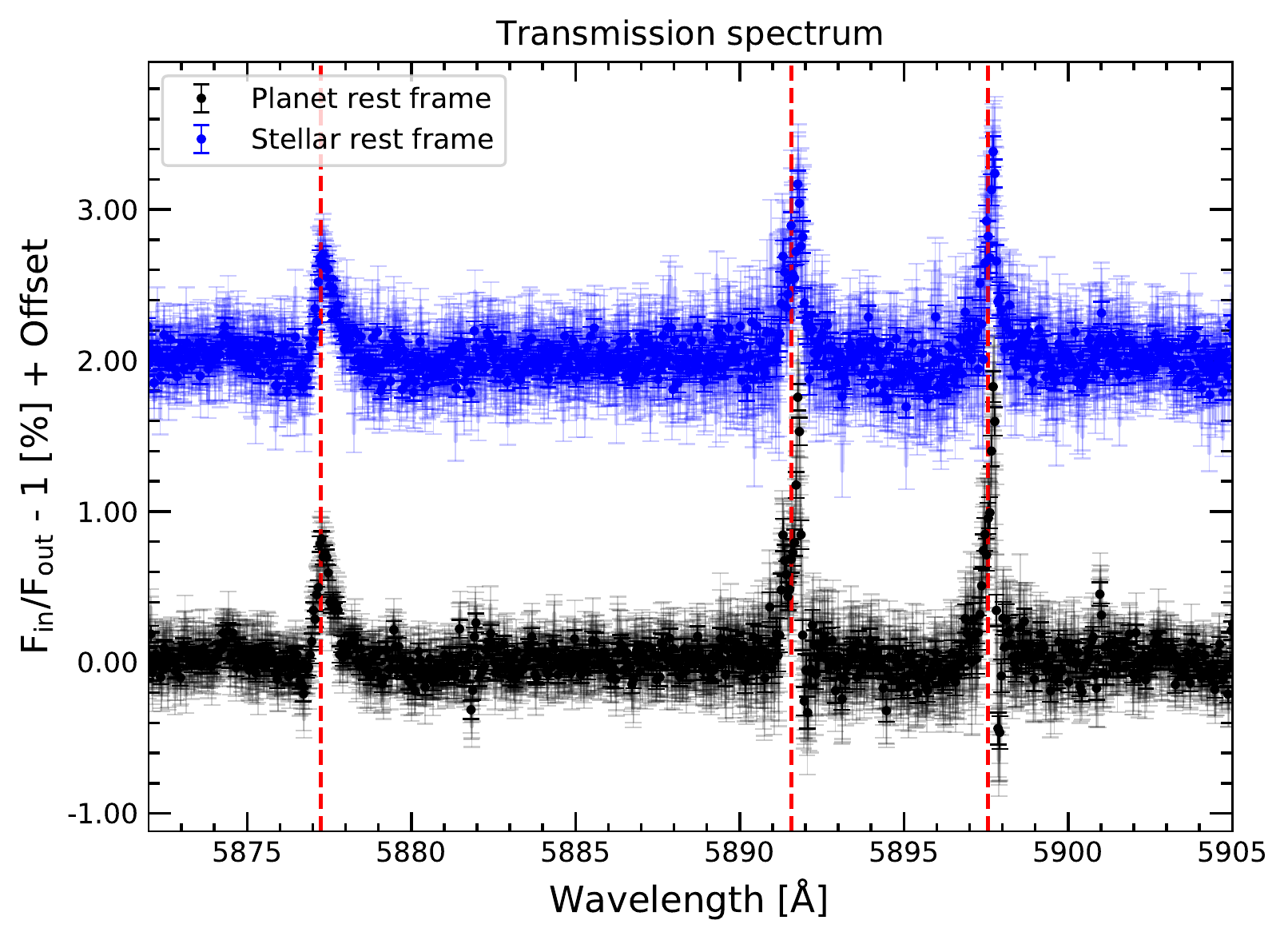}
	\caption{AU~Mic~b transmission spectrum around the \ion{Na}{i} doublet. The black dots show the original data with the respective error bars. The sinusoidal patter of the spectrum is due to ESPRESSO wiggle pattern, which we have not corrected here.}
	\label{fig:trans}
\end{figure}

AU~Mic~b's low density (0.6 that of Neptune; Plavchan et al 2020) could be explained by an extended H/He atmosphere with a large transit cross-section, and although the nominal equilibrium temperature of the planet is 600-800K, heating by XUV radiation from the host star (the $L_{x}/L_{bol}$ of AU Mic is $\sim$1000$\times$ that of the Sun) could maintain a temperature inversion and produce H-$\alpha$ absorption \citep{Yan2018}. Here, we probe the composition and structure of the upper atmosphere of AU Mic b via transmission spectroscopy to try to constrain models of H/He escape, which are fundamental to explaining aspects of the mass-radius relation and radius distribution of exoplanets \citep{Owen2013, Fulton17}.

To search for planetary atmospheric features, we use the 1D spectral product of the DRS software, and follow the methodology employed in \citet{Casasayas2020}. Prior to data analysis, the spectra are corrected for telluric absorptions using \texttt{Molecfit} \citep{Molecfit1, Molecfit2}. After that, the analysis steps are: i) moving the spectra to the stellar rest frame, using the system parameters from Plavchan et al (2020); ii) construct a master-out, high SNR, spectrum; iii) divide all spectra by the master-out and shift them to the planet's rest frame; iv) the combination of all the in-transit residual spectra \citep{Casasayas2019, Casasayas2020}; and v) Correction for the sinusoidal interference pattern in the the continuum that is sometimes seen in ESPRESSO (Casasayas et al, 2020b).

Unfortunately, stellar activity plays a big role in masking any possible planetary signal. Figure~\ref{fig:2danom} shows the 2D residual maps of our data series near the Na I and He I lines. Where one would expect a featureless map before and after the transit, strong structures in the data are seen both in and out of transit. The transmission spectrum for AU~Mic~b for this same spectral region is plotted in Figure~\ref{fig:trans}, where all three spectral lines are detected in emission, rather than in absorption if they were planetary signals in origin. The reason is the strong stellar flare that occurred during our observations near the transit egress, which lead to enhanced stellar chromospheric emission during transit. Thus, any possible planetary absorption will be masked by this emission. We tried the approach to discard data that are affected by flares, but nearly all the spectra are affected at a different levels, and we are quickly left with no out-of-transit observations (this is also true for the analysis in Section~\ref{sec:cc}).   

To illustrate the effects of stellar activity, we have focused on the He I line at 5877.2 \AA, and we built a transit light curve centered around this lines. To build the light curve, we again followed the methodology given in \citet{Casasayas2020}, integrating the line over a $1.5-\AA$ width interval. The light curve is shown in Figure~\ref{fig:indicators} and it traces the up to three flares that occur during our observations. The light curve of other activity-affected lines such as $H_{\alpha}$, the Na I doublet, or the Ca H\&K, correlate well with the He I (not shown), but being this a weaker and narrower spectral features, it is easier to construct a high SNR light curve. 

We have explored the full spectral range covered by ESPRESSO in search for detectable spectral absorption features, with null results. The spectral regions near the $Na~ I$, $K~ I$, $H_{\alpha}$, $Fe$ or $Mg$ most prominent absorption lines, are always dominated by the effects of stellar activity. Thus, we have to conclude that transmission spectroscopy for AU~Mic~b, and for young forming planets around active young stars in general, is going to remain very challenging if not impossible for the near future.


\subsection{Cross-Correlation studies}
\label{sec:cc}

\begin{figure}
	\centering
	\includegraphics[width=\columnwidth]{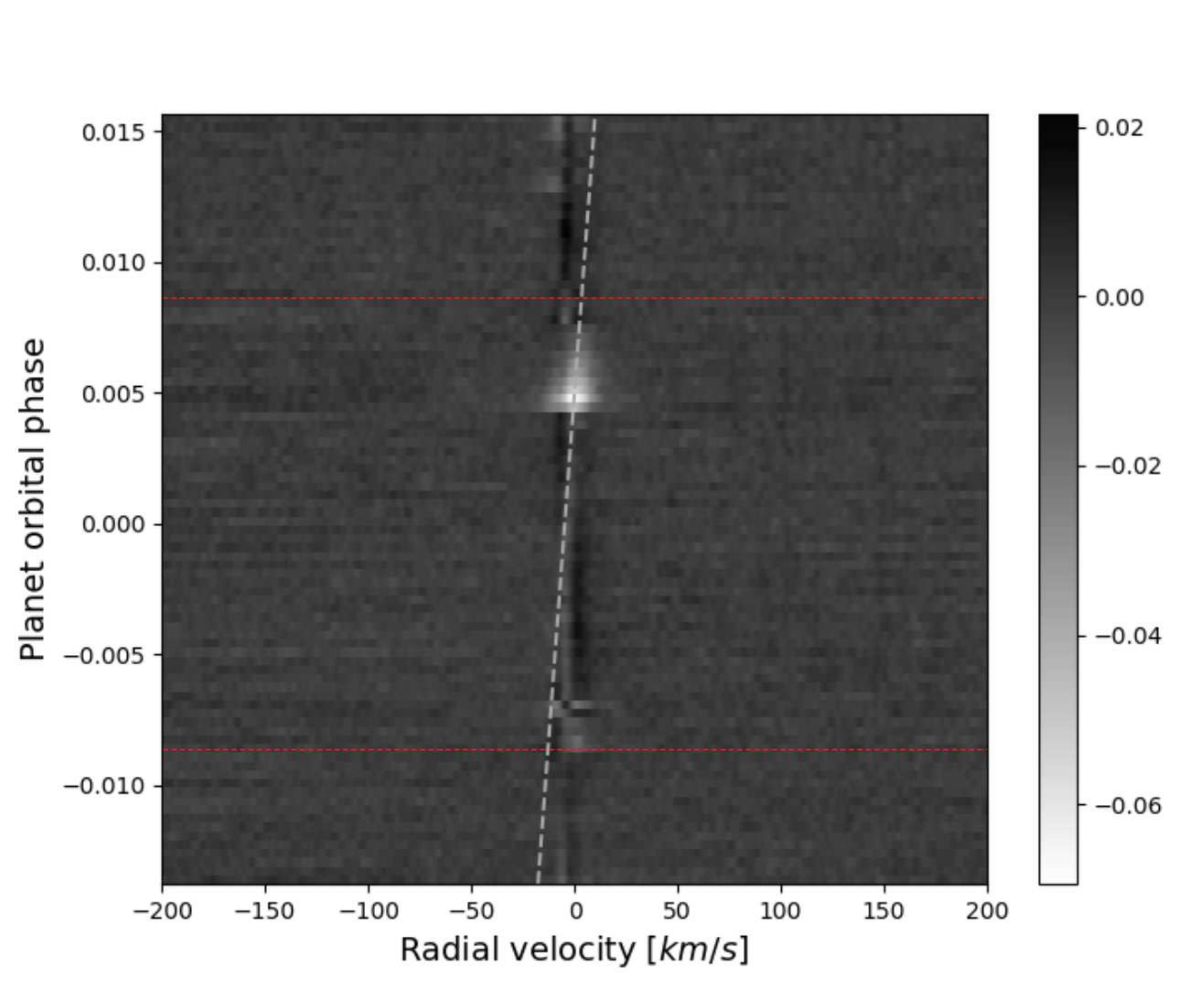}
	\includegraphics[width=\columnwidth]{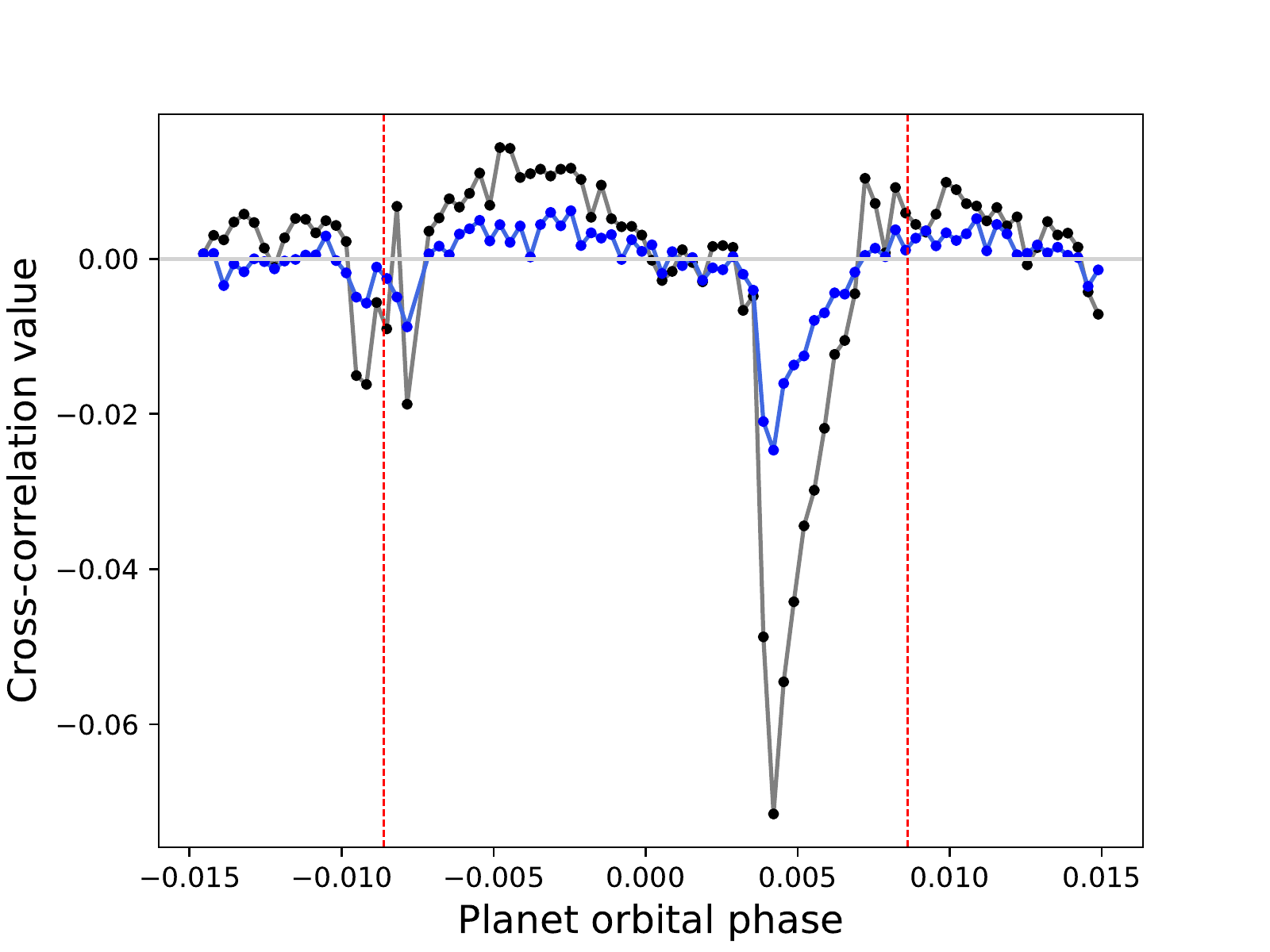}
	\caption{Top: Cross-correlation residuals map of $Fe I$ during the observations. The horizontal red dashed lines show first (T1) and last (T4) transit contacts and the white dashed lines show the expected planetary trail. Bottom:  A plot of the  cross-correlation values at 0 $km/s$ vs time for $Fe I$ (black), i.e. a vertical cut on the top panel. Also shown is the cross-correlation evolution but for the $Ca I$ (blue), showing a similar time evolution.}
	\label{fig:ccftime}
\end{figure}

Finally, we used the cross-correlation technique, following \citet{Stangret2020}, to search for atmospheric signatures in AU~Mic~b. We start from the \texttt{Molecfit} telluric-corrected data, and we removed outliers in the spectrum by studying the time evolution of each pixel, then we normalised each order of the spectrum and mask the sky emission lines as well as strong telluric absorption lines. Finally, we applied SYSREM \citep{Mazeh2007} to remove time- and wavelength-dependent trends in the spectral matrix.  
To calculate the high-resolution models used in the cross-correlation, we use petitRADTRANS \citep{Molliere2019}. For each studied individual atmospheric species we cross-correlated a model with residuals using radial velocities range +-200 km/s in steps of 0.5 km/s, and we added all the orders. As with the transmission spectroscopy technique, we find that the stellar variability prevents us from exploring the planetary atmosphere. Our cross-correlation residual maps for Fe I and similar atomic/ionized species (see Figure~\ref{fig:ccftime}) are completely dominated by stellar chromospheric emission.

\section{Conclusions}
\label{sec:Concl}

The proximity, brightness, youth, and the presence of a debris disk in AU~Mic mean that it will quickly become a key system for studies of exoplanet formation and atmospheric evolution. Particularly interesting is the fact that at least one of its planets, AU~Mic~b, transits the star offering a unique opportunity to further constrain the physical properties of the system. 

Following the recent discovery of AU~Mic~b, here we have conducted observations with ESPRESSO to measure the spin-orbit angle of the planets. We employed different data reduction and analysis techniques (Rossiter-McLaughlin effect measurement and doppler tomography) to deal with the activity-affected data, which all give consistent results. AU~Mic~b, seems to be well aligned with the rotation plane of its host star. This may not necessarily mean that the planet’s orbit is aligned with the current debris disk since there is a degeneracy in the angles on the sky.  However, this seems to be the case for AU Mic  \citep{Greaves2014, Watson2011}. This means that the formation and migration of the planets of the AU Mic system occurred within the disk. AU Mic~b is the third young ($<100 My$) small planet with a published measurement of its spin-orbit alignment, after DS Tuc b \citep{Zhou-20} and HD 63433 b \citep{Mann-20}. Noticeably, all three planets show prograde and aligned orbits.


We have also explored the full spectral range covered by ESPRESSO in search for detectable spectral absorption features, with null results. The spectral region near the $Na I$, $K I$, $H_{\alpha}$, $He I$, $Fe$ or $Mg$ most prominent absorption lines is always dominated by the effects of stellar activity. Thus, we have to conclude that transmission spectroscopy for recently formed planets around active young stars is going to remain very challenging if not impossible for the near future, at least at optical wavelengths.

\begin{acknowledgements}

This work is partly financed by the Spanish Ministry of Economics and Competitiveness through projects  PGC2018-098153-B-C31 and ESP2016-80435-C2-2-R.
This work was supported by JSPS KAKENHI Grant Number JP19K14783.

\end{acknowledgements}

\bibliographystyle{aa.bst} 
\bibliography{biblio.bib}

\begin{thebibliography}{63}
\expandafter\ifx\csname natexlab\endcsname\relax\def\natexlab#1{#1}\fi

\bibitem[{{Aigrain} {et~al.}(2016){Aigrain}, {Parviainen}, \&
  {Pope}}]{Aigrain-16}
{Aigrain}, S., {Parviainen}, H., \& {Pope}, B.~J.~S. 2016, \mnras, 459, 2408

\bibitem[{{Akinsanmi} {et~al.}(2018){Akinsanmi}, {Oshagh}, {Santos}, \&
  {Barros}}]{Akinsanmi-18}
{Akinsanmi}, B., {Oshagh}, M., {Santos}, N.~C., \& {Barros}, S.~C.~C. 2018,
  \aap, 609, A21

\bibitem[{{Albrecht} {et~al.}(2012){Albrecht}, {Winn}, {Butler}, {Crane},
  {Shectman}, {Thompson}, {Hirano}, \& {Wittenmyer}}]{Albrecht-12b}
{Albrecht}, S., {Winn}, J.~N., {Butler}, R.~P., {et~al.} 2012, \apj, 744, 189

\bibitem[{{Baruteau} {et~al.}(2016){Baruteau}, {Bai}, {Mordasini}, \&
  {Molli{\`e}re}}]{Baruteau2016}
{Baruteau}, C., {Bai}, X., {Mordasini}, C., \& {Molli{\`e}re}, P. 2016, \ssr,
  205, 77

\bibitem[{{Boccaletti} {et~al.}(2018){Boccaletti}, {Sezestre}, {Lagrange},
  {Th{\'e}bault}, {Gratton}, {Langlois}, {Thalmann}, {Janson}, {Delorme},
  {Augereau}, {Schneider}, {Milli}, {Grady}, {Debes}, {Kral}, {Olofsson},
  {Carson}, {Maire}, {Henning}, {Wisniewski}, {Schlieder}, {Dominik},
  {Desidera}, {Ginski}, {Hines}, {M{\'e}nard}, {Mouillet}, {Pawellek}, {Vigan},
  {Lagadec}, {Avenhaus}, {Beuzit}, {Biller}, {Bonavita}, {Bonnefoy},
  {Brandner}, {Cantalloube}, {Chauvin}, {Cheetham}, {Cudel}, {Gry}, {Daemgen},
  {Feldt}, {Galicher}, {Girard}, {Hagelberg}, {Janin-Potiron}, {Kasper}, {Le
  Coroller}, {Mesa}, {Peretti}, {Perrot}, {Samland}, {Sissa}, {Wildi}, {Zurlo},
  {Rochat}, {Stadler}, {Gluck}, {Orign{\'e}}, {Llored}, {Baudoz}, {Rousset},
  {Martinez}, \& {Rigal}}]{Boccaletti2018}
{Boccaletti}, A., {Sezestre}, E., {Lagrange}, A.~M., {et~al.} 2018, \aap, 614,
  A52

\bibitem[{{Boldt} {et~al.}(2020){Boldt}, {Oshagh}, {Dreizler}, {Mallonn},
  {Santos}, {Claret}, {Reiners}, \& {Sedaghati}}]{Boldt-20}
{Boldt}, S., {Oshagh}, M., {Dreizler}, S., {et~al.} 2020, \aap, 635, A123

\bibitem[{{Bou{\'e}} {et~al.}(2013){Bou{\'e}}, {Montalto}, {Boisse}, {Oshagh},
  \& {Santos}}]{Boue-12}
{Bou{\'e}}, G., {Montalto}, M., {Boisse}, I., {Oshagh}, M., \& {Santos}, N.~C.
  2013, \aap, 550, A53

\bibitem[{{Brown} {et~al.}(2017){Brown}, {Triaud}, {Doyle}, {Gillon}, {Lendl},
  {Anderson}, {Collier Cameron}, {H{\'e}brard}, {Hellier}, {Lovis}, {Maxted},
  {Pepe}, {Pollacco}, {Queloz}, \& {Smalley}}]{Brown-17}
{Brown}, D.~J.~A., {Triaud}, A.~H.~M.~J., {Doyle}, A.~P., {et~al.} 2017,
  \mnras, 464, 810

\bibitem[{{Butler} {et~al.}(1996){Butler}, {Marcy}, {Williams}, {McCarthy},
  {Dosanjh}, \& {Vogt}}]{Butler-96}
{Butler}, R.~P., {Marcy}, G.~W., {Williams}, E., {et~al.} 1996, \pasp, 108, 500

\bibitem[{{Casasayas-Barris} {et~al.}(2019){Casasayas-Barris}, {Pall{\'e}},
  {Yan}, {Chen}, {Kohl}, {Stangret}, {Parviainen}, {Helling}, {Watanabe},
  {Czesla}, {Fukui}, {Monta{\~n}{\'e}s-Rodr{\'\i}guez}, {Nagel}, {Narita},
  {Nortmann}, {Nowak}, {Schmitt}, \& {Zapatero Osorio}}]{Casasayas2019}
{Casasayas-Barris}, N., {Pall{\'e}}, E., {Yan}, F., {et~al.} 2019, \aap, 628,
  A9

\bibitem[{{Casasayas-Barris} {et~al.}(2020){Casasayas-Barris}, {Palle}, {Yan},
  {Chen}, {Luque}, {Stangret}, {Nagel}, {Zechmeister}, {Oshagh},
  {Sanz-Forcada}, {Nortmann}, {Alonso-Floriano}, {Amado}, {Caballero},
  {Czesla}, {Khalafinejad}, {Lopez-Puertas}, {Lopez-Santiago},
  {Molaverdikhani}, {Montes}, {Quirrenbach}, {Reiners}, {Ribas},
  {Sanchez-Lopez}, \& {Zapatero Osorio}}]{Casasayas2020}
{Casasayas-Barris}, N., {Palle}, E., {Yan}, F., {et~al.} 2020, arXiv e-prints,
  arXiv:2002.10595

\bibitem[{{Cegla} {et~al.}(2016{\natexlab{a}}){Cegla}, {Lovis}, {Bourrier},
  {Beeck}, {Watson}, \& {Pepe}}]{Cegla2016}
{Cegla}, H.~M., {Lovis}, C., {Bourrier}, V., {et~al.} 2016{\natexlab{a}}, \aap,
  588, A127

\bibitem[{{Cegla} {et~al.}(2016{\natexlab{b}}){Cegla}, {Oshagh}, {Watson},
  {Figueira}, {Santos}, \& {Shelyag}}]{Cegla-16}
{Cegla}, H.~M., {Oshagh}, M., {Watson}, C.~A., {et~al.} 2016{\natexlab{b}},
  \apj, 819, 67

\bibitem[{{Collier Cameron} {et~al.}(2010){Collier Cameron}, {Bruce}, {Miller},
  {Triaud}, \& {Queloz}}]{Collier2010}
{Collier Cameron}, A., {Bruce}, V.~A., {Miller}, G.~R.~M., {Triaud},
  A.~H.~M.~J., \& {Queloz}, D. 2010, \mnras, 403, 151

\bibitem[{{Csizmadia} {et~al.}(2013){Csizmadia}, {Pasternacki}, {Dreyer},
  {Cabrera}, {Erikson}, \& {Rauer}}]{Csizmadia2013}
{Csizmadia}, S., {Pasternacki}, T., {Dreyer}, C., {et~al.} 2013, \aap, 549, A9

\bibitem[{{Daley} {et~al.}(2019){Daley}, {Hughes}, {Carter}, {Flaherty},
  {Lambros}, {Pan}, {Schlichting}, {Chiang}, {Wyatt}, {Wilner}, {Andrews}, \&
  {Carpenter}}]{Daley2019}
{Daley}, C., {Hughes}, A.~M., {Carter}, E.~S., {et~al.} 2019, \apj, 875, 87

\bibitem[{{de Mooij} {et~al.}(2017){de Mooij}, {Watson}, \&
  {Kenworthy}}]{deMooij-17}
{de Mooij}, E.~J.~W., {Watson}, C.~A., \& {Kenworthy}, M.~A. 2017, \mnras, 472,
  2713

\bibitem[{{Faria} {et~al.}(2016){Faria}, {Haywood}, {Brewer}, {Figueira},
  {Oshagh}, {Santerne}, \& {Santos}}]{Faria-16}
{Faria}, J.~P., {Haywood}, R.~D., {Brewer}, B.~J., {et~al.} 2016, \aap, 588,
  A31

\bibitem[{{Feroz} {et~al.}(2009){Feroz}, {Hobson}, \& {Bridges}}]{Feroz-09}
{Feroz}, F., {Hobson}, M.~P., \& {Bridges}, M. 2009, \mnras, 398, 1601

\bibitem[{{Foreman-Mackey} {et~al.}(2017){Foreman-Mackey}, {Agol}, {Angus}, \&
  {Ambikasaran}}]{Foreman-Mackey-17}
{Foreman-Mackey}, D., {Agol}, E., {Angus}, R., \& {Ambikasaran}, S. 2017, ArXiv

\bibitem[{{Foreman-Mackey} {et~al.}(2013){Foreman-Mackey}, {Hogg}, {Lang}, \&
  {Goodman}}]{Foreman-Mackey-13}
{Foreman-Mackey}, D., {Hogg}, D.~W., {Lang}, D., \& {Goodman}, J. 2013, \pasp,
  125, 306

\bibitem[{{Fulton} {et~al.}(2017){Fulton}, {Petigura}, {Howard}, {Isaacson},
  {Marcy}, {Cargile}, {Hebb}, {Weiss}, {Johnson}, {Morton}, {Sinukoff},
  {Crossfield}, \& {Hirsch}}]{Fulton17}
{Fulton}, B.~J., {Petigura}, E.~A., {Howard}, A.~W., {et~al.} 2017, \aj, 154,
  109

\bibitem[{{Greaves} {et~al.}(2014){Greaves}, {Kennedy}, {Thureau}, {Eiroa},
  {Marshall}, {Maldonado}, {Matthews}, {Olofsson}, {Barlow},
  {Moro-Mart{\'\i}n}, {Sibthorpe}, {Absil}, {Ardila}, {Booth},
  {Broekhoven-Fiene}, {Brown}, {Collier Cameron}, {del Burgo}, {Di Francesco},
  {Eisl{\"o}ffel}, {Duch{\^e}ne}, {Ertel}, {Holland}, {Horner}, {Kalas},
  {Kavelaars}, {Lestrade}, {Vican}, {Wilner}, {Wolf}, \& {Wyatt}}]{Greaves2014}
{Greaves}, J.~S., {Kennedy}, G.~M., {Thureau}, N., {et~al.} 2014, \mnras, 438,
  L31

\bibitem[{{Haywood} {et~al.}(2014){Haywood}, {Collier Cameron}, {Queloz},
  {Barros}, {Deleuil}, {Fares}, {Gillon}, {Lanza}, {Lovis}, {Moutou}, {Pepe},
  {Pollacco}, {Santerne}, {S{\'e}gransan}, \& {Unruh}}]{Haywood-14}
{Haywood}, R.~D., {Collier Cameron}, A., {Queloz}, D., {et~al.} 2014, \mnras,
  443, 2517

\bibitem[{{Hirano}(2014)}]{Hirano-14}
{Hirano}, T. 2014, {Measurement of Spin-Orbit Angles for Transiting Systems:
  Toward an Understanding of the Migration History of Planets.}

\bibitem[{{Hirano} {et~al.}(2020){Hirano}, {Gaidos}, {Winn}, {Dai}, {Fukui},
  {Kuzuhara}, {Kotani}, {Tamura}, {Hjorth}, {Albrecht}, {Huber}, {Bolmont},
  {Harakawa}, {Hodapp}, {Ishizuka}, {Jacobson}, {Konishi}, {Kudo}, {Kurokawa},
  {Nishikawa}, {Omiya}, {Serizawa}, {Ueda}, \& {Weiss}}]{2020ApJ...890L..27H}
{Hirano}, T., {Gaidos}, E., {Winn}, J.~N., {et~al.} 2020, \apjl, 890, L27

\bibitem[{{Hirano} {et~al.}(2011){Hirano}, {Suto}, {Winn}, {Taruya}, {Narita},
  {Albrecht}, \& {Sato}}]{Hirano-11}
{Hirano}, T., {Suto}, Y., {Winn}, J.~N., {et~al.} 2011, \apj, 742, 69

\bibitem[{{Holt}(1893)}]{Holt-1893}
{Holt}, J.~R. 1893, Astronomy and Astro-Physics (formerly The Sidereal
  Messenger), 12, 646

\bibitem[{{Kausch} {et~al.}(2015){Kausch}, {Noll}, {Smette}, {Kimeswenger},
  {Barden}, {Szyszka}, {Jones}, {Sana}, {Horst}, \& {Kerber}}]{Molecfit2}
{Kausch}, W., {Noll}, S., {Smette}, A., {et~al.} 2015, \aap, 576, A78

\bibitem[{{Kley} \& {Nelson}(2012)}]{Kley2012}
{Kley}, W. \& {Nelson}, R.~P. 2012, \araa, 50, 211

\bibitem[{{Lin} {et~al.}(1996){Lin}, {Bodenheimer}, \& {Richardson}}]{Lin1996}
{Lin}, D.~N.~C., {Bodenheimer}, P., \& {Richardson}, D.~C. 1996, \nat, 380, 606

\bibitem[{{Mann} {et~al.}(2016){Mann}, {Gaidos}, {Mace}, {Johnson}, {Bowler},
  {LaCourse}, {Jacobs}, {Vanderburg}, {Kraus}, {Kaplan}, \& {Jaffe}}]{Mann2016}
{Mann}, A.~W., {Gaidos}, E., {Mace}, G.~N., {et~al.} 2016, \apj, 818, 46

\bibitem[{{Mann} {et~al.}(2020){Mann}, {Johnson}, {Vanderburg}, {Kraus},
  {Rizzuto}, {Wood}, {Bush}, {Rockcliffe}, {Newton}, {Latham}, {Mamajek},
  {Zhou}, {Quinn}, {Thao}, {Benatti}, {Cosentino}, {Desidera}, {Harutyunyan},
  {Lovis}, {Mortier}, {Pepe}, {Poretti}, {Wilson}, {Kristiansen}, {Gagliano},
  {Jacobs}, {LaCourse}, {Omohundro}, {Schwengeler}, {Kane}, {Hill}, {Rabus},
  {Esquerdo}, {Berlind}, {Collins}, {Murawski}, {Aitken}, {Hazam Sallam},
  {Massey}, {Ricker}, {Vanderspek}, {Seager}, {Winn}, {Jenkins}, {Barclay},
  {Caldwell}, {Dragomir}, {Doty}, {Glidden}, {Tenenbaum}, {Torres}, {Twicken},
  \& {Villanueva}}]{Mann-20}
{Mann}, A.~W., {Johnson}, M.~C., {Vanderburg}, A., {et~al.} 2020, arXiv
  e-prints, arXiv:2005.00047

\bibitem[{{Mazeh} {et~al.}(2007){Mazeh}, {Tamuz}, \& {Zucker}}]{Mazeh2007}
{Mazeh}, T., {Tamuz}, O., \& {Zucker}, S. 2007, Astronomical Society of the
  Pacific Conference Series, Vol. 366, {The Sys-Rem Detrending Algorithm:
  Implementation and Testing}, ed. C.~{Afonso}, D.~{Weldrake}, \& T.~{Henning},
  119

\bibitem[{{McLaughlin}(1924)}]{McLaughlin-24}
{McLaughlin}, D.~B. 1924, \apj, 60, 22

\bibitem[{{Molli{\`e}re} {et~al.}(2019){Molli{\`e}re}, {Wardenier}, {van
  Boekel}, {Henning}, {Molaverdikhani}, \& {Snellen}}]{Molliere2019}
{Molli{\`e}re}, P., {Wardenier}, J.~P., {van Boekel}, R., {et~al.} 2019, \aap,
  627, A67

\bibitem[{{Ohta} {et~al.}(2005){Ohta}, {Taruya}, \& {Suto}}]{Ohta-05}
{Ohta}, Y., {Taruya}, A., \& {Suto}, Y. 2005, \apj, 622, 1118

\bibitem[{{Oshagh} {et~al.}(2013){Oshagh}, {Bou{\'e}}, {Figueira}, {Santos}, \&
  {Haghighipour}}]{Oshagh-13}
{Oshagh}, M., {Bou{\'e}}, G., {Figueira}, P., {Santos}, N.~C., \&
  {Haghighipour}, N. 2013, \aap, 558, A65

\bibitem[{{Oshagh} {et~al.}(2016){Oshagh}, {Dreizler}, {Santos}, {Figueira}, \&
  {Reiners}}]{Oshagh-16}
{Oshagh}, M., {Dreizler}, S., {Santos}, N.~C., {Figueira}, P., \& {Reiners}, A.
  2016, \aap, 593, A25

\bibitem[{{Oshagh} {et~al.}(2018){Oshagh}, {Triaud}, {Burdanov}, {Figueira},
  {Reiners}, {Santos}, {Faria}, {Boue}, {D{\'{\i}}az}, {Dreizler}, {Boldt},
  {Delrez}, {Ducrot}, {Gillon}, {Guzman Mesa}, {Jehin}, {Khalafinejad}, {Kohl},
  {Serrano}, \& {Udry}}]{Oshagh-18}
{Oshagh}, M., {Triaud}, A.~H.~M.~J., {Burdanov}, A., {et~al.} 2018, \aap, 619,
  A150

\bibitem[{{Owen} \& {Wu}(2013)}]{Owen2013}
{Owen}, J.~E. \& {Wu}, Y. 2013, \apj, 775, 105

\bibitem[{{Pepe} {et~al.}(2002){Pepe}, {Mayor}, {Galland}, {Naef}, {Queloz},
  {Santos}, {Udry}, \& {Burnet}}]{Pepe-02}
{Pepe}, F., {Mayor}, M., {Galland}, F., {et~al.} 2002, \aap, 388, 632

\bibitem[{{Prato} {et~al.}(2008){Prato}, {Huerta}, {Johns-Krull}, {Mahmud},
  {Jaffe}, \& {Hartigan}}]{Prato2008}
{Prato}, L., {Huerta}, M., {Johns-Krull}, C.~M., {et~al.} 2008, \apjl, 687,
  L103

\bibitem[{{Rasmussen} \& {Williams}(2006)}]{Rasmussen-06}
{Rasmussen}, C.~E. \& {Williams}, C. K.~I. 2006, {Gaussian Processes for
  Machine Learning}

\bibitem[{Ricker {et~al.}(2014)Ricker, Winn, Vanderspek, Latham, Bakos, Bean,
  Berta-Thompson, Brown, Buchhave, Butler, Butler, Chaplin, Charbonneau,
  Christensen-Dalsgaard, Clampin, Deming, Doty, {De Lee}, Dressing, Dunham,
  Endl, Fressin, Ge, Henning, Holman, Howard, Ida, Jenkins, Jernigan, Johnson,
  Kaltenegger, Kawai, Kjeldsen, Laughlin, Levine, Lin, Lissauer, MacQueen,
  Marcy, McCullough, Morton, Narita, Paegert, Palle, Pepe, Pepper, Quirrenbach,
  Rinehart, Sasselov, Sato, Seager, Sozzetti, Stassun, Sullivan, Szentgyorgyi,
  Torres, Udry, \& Villasenor}]{Ricker2014}
Ricker, G.~R., Winn, J.~N., Vanderspek, R., {et~al.} 2014, 914320

\bibitem[{{Rossiter}(1924)}]{Rossiter-24}
{Rossiter}, R.~A. 1924, \apj, 60, 15

\bibitem[{{Serrano} {et~al.}(2018){Serrano}, {Barros}, {Oshagh}, {Santos},
  {Faria}, {Demangeon}, {Sousa}, \& {Lendl}}]{Serrano-18}
{Serrano}, L.~M., {Barros}, S.~C.~C., {Oshagh}, M., {et~al.} 2018, \aap, 611,
  A8

\bibitem[{{Serrano} {et~al.}(2020){Serrano}, {Oshagh}, {Cegla}, {Barros},
  {Santos}, {Faria}, \& {Akinsanmi}}]{Serrano-20}
{Serrano}, L.~M., {Oshagh}, M., {Cegla}, H.~M., {et~al.} 2020, \mnras, 493,
  5928

\bibitem[{{Shporer} \& {Brown}(2011)}]{Shporer-11}
{Shporer}, A. \& {Brown}, T. 2011, \apj, 733, 30

\bibitem[{{Smette} {et~al.}(2015){Smette}, {Sana}, {Noll}, {Horst}, {Kausch},
  {Kimeswenger}, {Barden}, {Szyszka}, {Jones}, {Gallenne}, {Vinther},
  {Ballester}, \& {Taylor}}]{Molecfit1}
{Smette}, A., {Sana}, H., {Noll}, S., {et~al.} 2015, \aap, 576, A77

\bibitem[{{Speagle}(2020)}]{Speagle-20}
{Speagle}, J.~S. 2020, \mnras, 493, 3132

\bibitem[{{Stangret} {et~al.}(2020){Stangret}, {Casasayas-Barris}, {Pall{\'e}},
  {Yan}, {S{\'a}nchez-L{\'o}pez}, \& {L{\'o}pez-Puertas}}]{Stangret2020}
{Stangret}, M., {Casasayas-Barris}, N., {Pall{\'e}}, E., {et~al.} 2020, arXiv
  e-prints, arXiv:2003.04650

\bibitem[{{Strachan} \& {Anglada-Escud{\'e}}(2017)}]{Strachan2017}
{Strachan}, J. B.~P. \& {Anglada-Escud{\'e}}, G. 2017, \mnras, 472, 3467

\bibitem[{{Triaud}(2017)}]{Triaud-18}
{Triaud}, A.~H.~M.~J. 2017, {The Rossiter-McLaughlin Effect in Exoplanet
  Research} (Springer Living Reference Work), 2

\bibitem[{{Trotta}(2008)}]{Trotta-08}
{Trotta}, R. 2008, Contemporary Physics, 49, 71

\bibitem[{{Walkowicz} {et~al.}(2008){Walkowicz}, {Johns-Krull}, \&
  {Hawley}}]{Walkowicz2008}
{Walkowicz}, L.~M., {Johns-Krull}, C.~M., \& {Hawley}, S.~L. 2008, \apj, 677,
  593

\bibitem[{{Watson} {et~al.}(2019){Watson}, {de Mooij}, {Steeghs}, {Marsh},
  {Brogi}, {Gibson}, \& {Matthews}}]{Watson2019}
{Watson}, C.~A., {de Mooij}, E.~J.~W., {Steeghs}, D., {et~al.} 2019, \mnras,
  490, 1991

\bibitem[{{Watson} {et~al.}(2011){Watson}, {Littlefair}, {Diamond}, {Collier
  Cameron}, {Fitzsimmons}, {Simpson}, {Moulds}, \& {Pollacco}}]{Watson2011}
{Watson}, C.~A., {Littlefair}, S.~P., {Diamond}, C., {et~al.} 2011, \mnras,
  413, L71

\bibitem[{{Yan} {et~al.}(2015){Yan}, {Fosbury}, {Petr-Gotzens}, {Zhao}, \&
  {Pall{\'e}}}]{Yan-15}
{Yan}, F., {Fosbury}, R.~A.~E., {Petr-Gotzens}, M.~G., {Zhao}, G., \&
  {Pall{\'e}}, E. 2015, \aap, 574, A94

\bibitem[{{Yan} \& {Henning}(2018)}]{Yan2018}
{Yan}, F. \& {Henning}, T. 2018, Nature Astronomy, 2, 714

\bibitem[{{Zechmeister} {et~al.}(2018{\natexlab{a}}){Zechmeister}, {Reiners},
  {Amado}, {Azzaro}, {Bauer}, {B{\'e}jar}, {Caballero}, {Guenther}, {Hagen},
  {Jeffers}, {Kaminski}, {K{\"u}rster}, {Launhardt}, {Montes}, {Morales},
  {Quirrenbach}, {Reffert}, {Ribas}, {Seifert}, {Tal-Or}, \&
  {Wolthoff}}]{SERVAL}
{Zechmeister}, M., {Reiners}, A., {Amado}, P.~J., {et~al.} 2018{\natexlab{a}},
  \aap, 609, A12

\bibitem[{{Zechmeister} {et~al.}(2018{\natexlab{b}}){Zechmeister}, {Reiners},
  {Amado}, {Azzaro}, {Bauer}, {B{\'e}jar}, {Caballero}, {Guenther}, {Hagen},
  {Jeffers}, {Kaminski}, {K{\"u}rster}, {Launhardt}, {Montes}, {Morales},
  {Quirrenbach}, {Reffert}, {Ribas}, {Seifert}, {Tal-Or}, \&
  {Wolthoff}}]{Zechmeister-18}
{Zechmeister}, M., {Reiners}, A., {Amado}, P.~J., {et~al.} 2018{\natexlab{b}},
  \aap, 609, A12

\bibitem[{{Zhou} {et~al.}(2020){Zhou}, {Winn}, {Newton}, {Quinn}, {Rodriguez},
  {Mann}, {Rizzuto}, {Vand erburg}, {Huang}, {Latham}, {Teske}, {Wang},
  {Shectman}, {Butler}, {Crane}, {Thompson}, {Henry}, {Paredes}, {Jao},
  {James}, \& {Hinojosa}}]{Zhou-20}
{Zhou}, G., {Winn}, J.~N., {Newton}, E.~R., {et~al.} 2020, \apjl, 892, L21

\end{thebibliography}

\appendix

\section{Corner plots}


\begin{figure}
	\centering
	\includegraphics[width=\columnwidth]{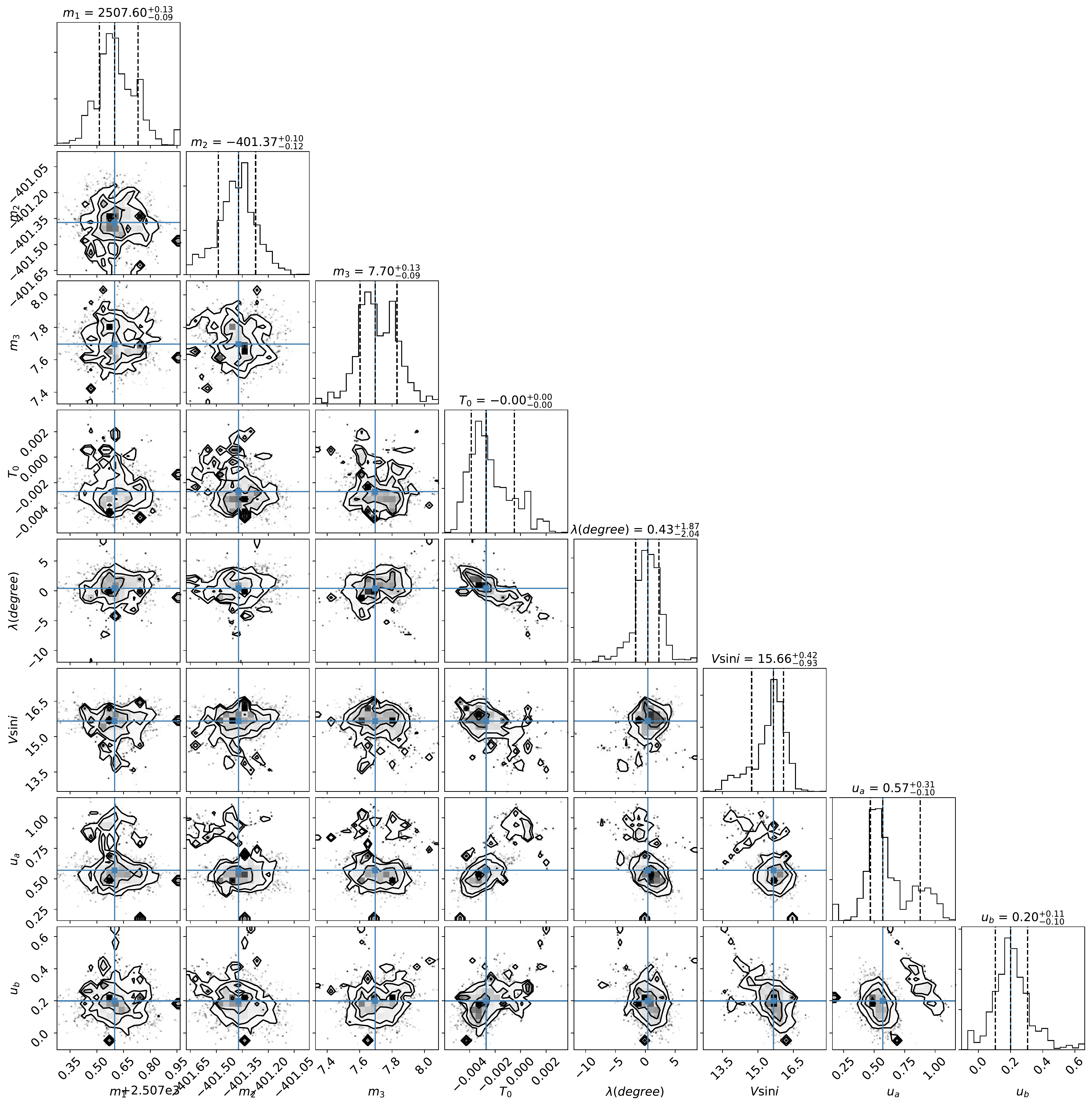}
	\caption{CAption }
	\label{fig:corner1}
\end{figure}

\begin{figure}
	\centering
	\includegraphics[width=\columnwidth]{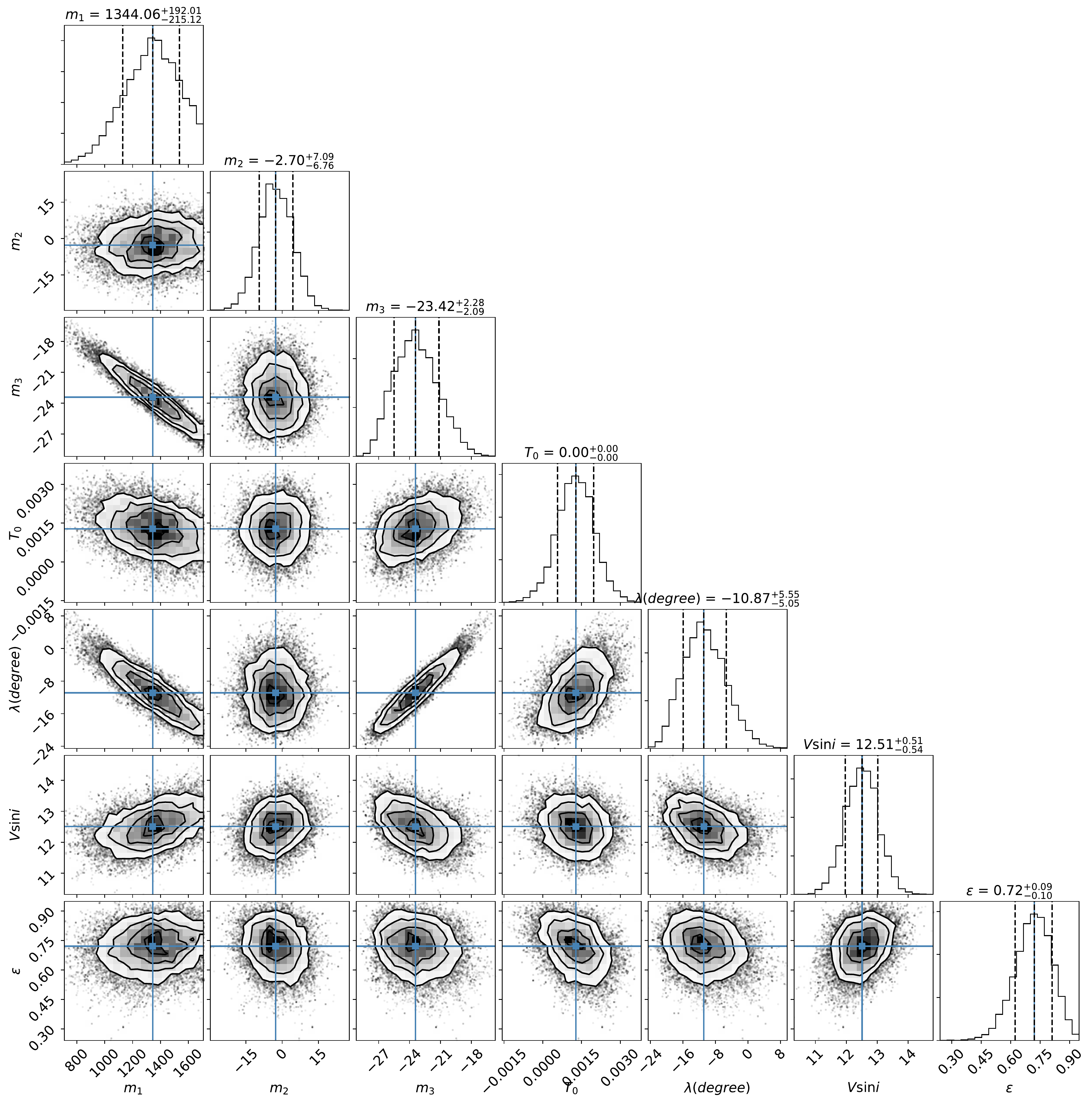}
	\caption{CAption }
	\label{fig:corner2}
\end{figure}

\begin{figure}
	\centering
	\includegraphics[width=\columnwidth]{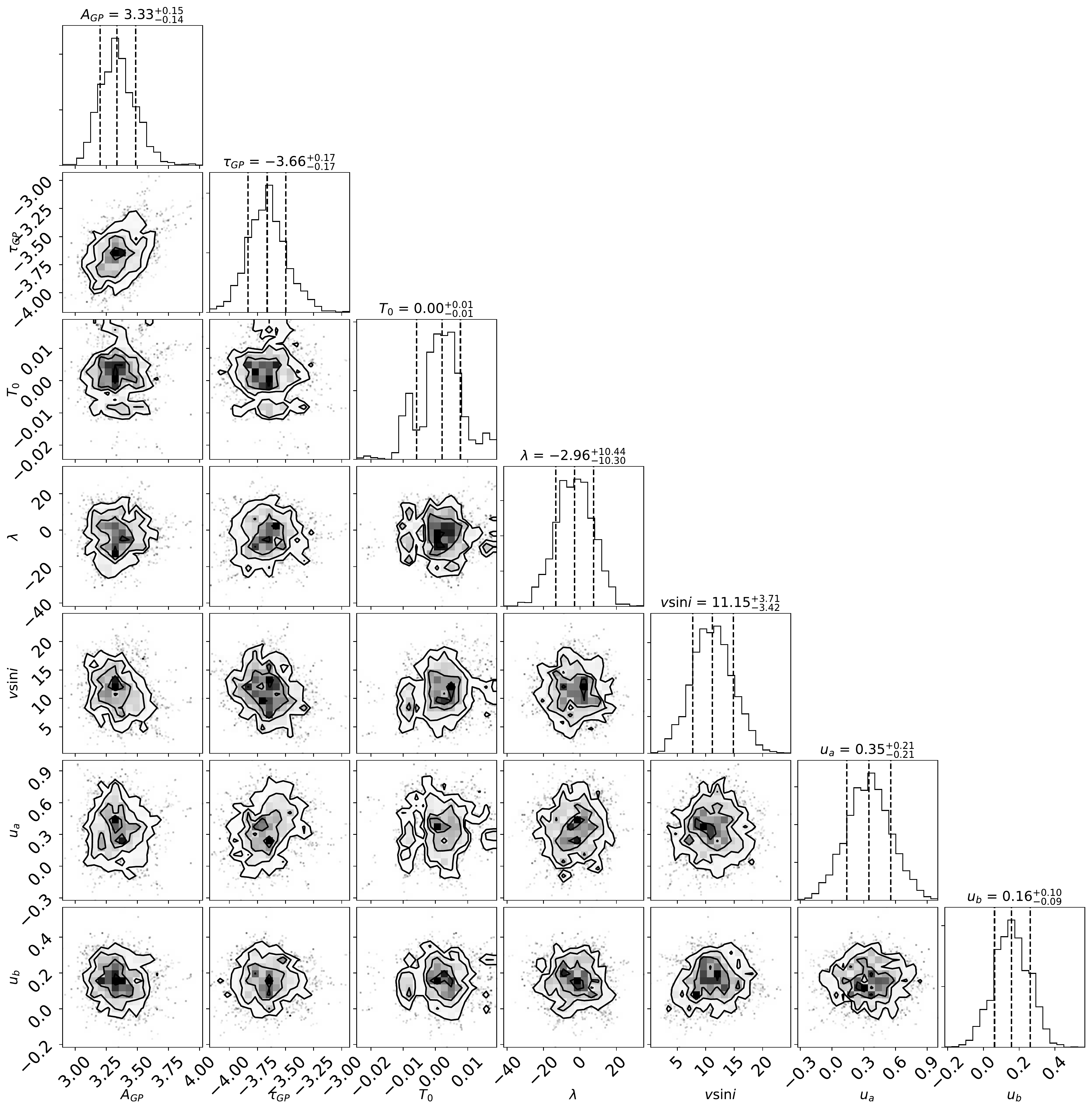}
	\caption{CAption }
	\label{fig:corner3}
\end{figure}

\begin{figure}
	\centering
	\includegraphics[width=\columnwidth]{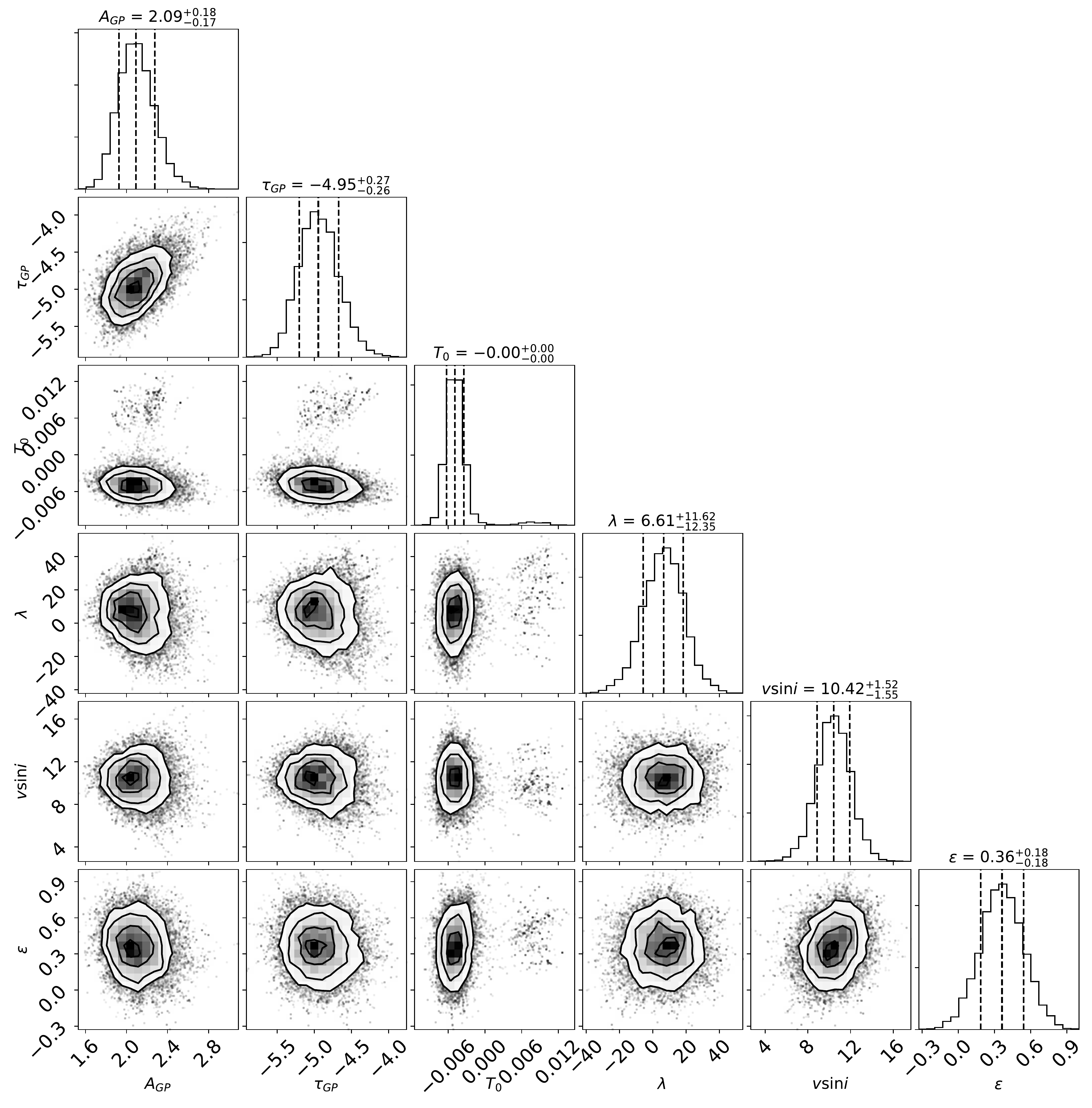}
	\caption{CAption }
	\label{fig:corner4}
\end{figure}

\end{document}